\documentclass[journal]{IEEEtran}
\usepackage{graphicx,subfigure,amsmath,amsfonts,amssymb,algorithm,algorithmic,cite,multirow,rotating,multicol}

\begin{document}

\title{Multi-View Broad Learning System\\ for Primate Oculomotor Decision Decoding}

\author{Zhenhua~Shi, Xiaomo~Chen, Changming~Zhao, He~He, Veit~Stuphorn and~Dongrui~Wu
\thanks{Z.~Shi, C.~Zhao, H.~He and D.~Wu are with the Ministry of Education Key Laboratory of Image Processing and Intelligent Control, School of Artificial Intelligence and Automation, Huazhong University of Science and Technology, Wuhan, China. Email: zhenhuashi@hust.edu.cn, cmzhao@hust.edu.cn, hehe91@hust.edu.cn, drwu@hust.edu.cn.}
\thanks{X.~Chen was with the Zanvyl Krieger Mind and Brain Institute, Johns Hopkins University School of Medicine. She is currently with the Department of Neurobiology, Stanford University, Stanford, CA, USA. Email: xiaomo@stanford.edu.}
\thanks{V.~Stuphorn is with the Department of Neuroscience, Johns Hopkins University, Baltimore, MD, USA. He is also with the Zanvyl Krieger Mind and Brain Institute, Johns Hopkins University School of Medicine. Email: veit@jhu.edu.}
\thanks{Z.~Shi and X.~Chen contributed equally to this work.}
\thanks{V.~Stuphorn and D.~Wu are the corresponding authors.}}

\maketitle

\begin{abstract}
Multi-view learning improves the learning performance by utilizing multi-view data: data collected from multiple sources, or feature sets extracted from the same data source. This approach is suitable for primate brain state decoding using cortical neural signals. This is because the complementary components of simultaneously recorded neural signals, local field potentials (LFPs) and action potentials (spikes), can be treated as two views. In this paper, we extended broad learning system (BLS), a recently proposed wide neural network architecture, from single-view learning to multi-view learning, and validated its performance in decoding monkeys' oculomotor decision from medial frontal LFPs and spikes. We demonstrated that medial frontal LFPs and spikes in non-human primate do contain complementary information about the oculomotor decision, and that the proposed multi-view BLS is a more effective approach for decoding the oculomotor decision than several classical and state-of-the-art single-view and multi-view learning approaches.
\end{abstract}

\IEEEpeerreviewmaketitle

\section{Introduction}

\IEEEPARstart{M}{ulti}-view learning attempts to improve the learning performance by utilizing multi-view data, which can be collected from multiple data sources, or different feature sets extracted from the same data source. For example, in an invasive brain-machine interface (BMI) using electrodes~\cite{Stavisky2015high}, effective BMI cursor control can be achieved using action potentials (spikes), which are high-pass filtered neural signals, or local field potentials (LFPs), which are low-pass filtered neural signals measured from the same electrodes. The spikes and LFPs can represent two views of the same task.

%Many multi-view learning algorithms have been proposed in the literature, which can be roughly categorized into the following three groups \cite{xu2013survey}:
%\begin{enumerate}
%\item Co-training \cite{blum1998combining} and its variants (e.g., co-EM \cite{nigam2000analyzing}), where a separate learner is trained on each view, but the learners are also correlated in that their outputs are forced to be similar on the same validation samples.
%
%\item Multi-kernel learning such as semi-definite programming \cite{lanckriet2004learning} and quadratically constrained quadratic programming \cite{bach2004multiple}, in which multiple kernels corresponding to inputs from different representations are combined to integrate information from different views.
%
%\item Subspace learning such as canonical correlation analysis (CCA) \cite{hotelling1936relations}, in which a latent subspace shared by multiple views is assumed to exist and is estimated.
%\end{enumerate}

There have been a few studies on applying multi-view learning to human brain state decoding. Kandemir \emph{et al.}~\cite{Kandemir2014} combined multi-task learning and multi-view learning in decoding a user's affective state, by treating different types of physiological sensors (e.g., electroencephalography, electrocardiography, etc.) as different views. Pasupa and Szedmak \cite{Pasupa2017} used tensor-based multi-view learning to predict where people are looking in images (saliency prediction), by treating the image and the user's eye movement as two views. Spyrou \emph{et al.} \cite{Spyrou2018} used multi-view learning to integrate spatial, temporal, and frequency signatures of electroencephalography signals for interictal epileptic discharges classification. However, to our knowledge, no one has applied multi-view learning to non-human primate brain state decoding using invasive signals like LFPs and spikes (more details can be found in in Section~\ref{sect:related}).

A broad learning system (BLS) \cite{chen2018broad} is a flexible neural network, which can incrementally adjust the number of nodes for the best performance. It has achieved comparable performance, with much less computational cost, to deep learning approaches in two applications \cite{chen2018broad}. The main difference between a BLS and a deep learning model is that BLS improves the learning performance by increasing the width, instead of the depth, of the neural network. This paper proposes a multi-view BLS (MvBLS), which extends BLS from traditional single-view learning to multi-view learning, and applies it to monkey oculomotor decision decoding from both LFP and spike features. By using features from different views in generating the enhancement nodes, the proposed MvBLS can significantly outperform some classical and state-of-the-art single-view and multi-view learning approaches.

The main contributions of this paper are:
\begin{enumerate}
\item We propose three different MvBLS architectures, which have comparable performances but different computational cost.

\item We apply MvBLS to monkey oculomotor decision decoding using neural signals recorded in the medial frontal cortex, and demonstrate that it outperformed some classical and state-of-the-art single-view and multi-view learning approaches.

\item We verify through extensive experiments that combining LFP and spike features can improve the decoding performance in monkey oculomotor decision classification. This shows that, at least in this context, LFPs and spikes in the medial frontal cortex contain complementary information about oculomotor decisions.
\end{enumerate}

The remainder of this paper is organized as follows: Section~II introduces the single-view BLS and our proposed MvBLS. Section~III describes the neurophysiological dataset used in this work, and the experimental results. Section~IV presents some additional discussions. Finally, Section~V draws conclusions.

\section{BLS and MvBLS}

This section introduces a single-view BLS and our proposed MvBLS for multi-class classification.

\subsection{Broad Learning System (BLS)}\label{sect:BLS}

Single-layer feed-forward neural networks are universal approximators when the underlying function is continuous \cite{leshno1993multilayer}, and have been used in numerous applications. Random vector functional neural networks (RVFLNNs) accelerate single-layer feed-forward neural networks by randomly generating the weight matrix \cite{igelnik1995stochastic}. BLS is a further improvement of the RVFLNN.

In an RVFLNN, the input and output layers are directly connected. In a BLS, the input layer first passes through a feature extractor for dimensionality reduction and noise suppression. Due to the use of sparse auto-encoders, the extracted features are more diverse. This helps improve the generalization performance.

\begin{figure}[htpb] \centering
\setlength{\abovecaptionskip}{0.cm}
\setlength{\belowcaptionskip}{-0.cm}
\includegraphics[width=.6\linewidth,clip]{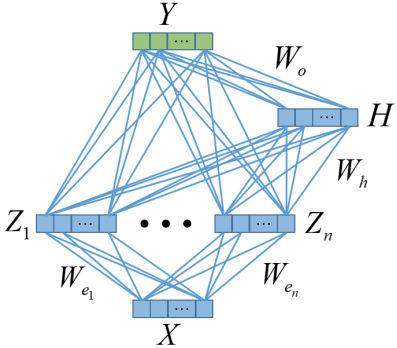}
\caption{The architecture of a BLS \cite{chen2018broad}. Diverse linear de-noised features $\mathbf{Z}$ are extracted from data $\mathbf{X}$, and further mapped into nonlinear features $\mathbf{H}$. Linear features $\mathbf{Z}$ and nonlinear features $\mathbf{H}$ are then concatenated to predict $\mathbf{Y}$. } \label{fig:BLS}
\end{figure}

Let $\mathbf{X}\in \mathbb{R}^{N\times M}$ be the data matrix, where $N$ is the number of observations, and $M$ the feature dimensionality. Let $\mathbf{Y}\in \mathbb{R}^{N\times C}$ be the one-hot coding matrix of the labels of $\mathbf{X}$, where $C$ is the number of classes. The architecture of a BLS is shown in Fig.~\ref{fig:BLS}. It first constructs feature nodes $\mathbf{Z}$ from $\mathbf{X}$, and then enhancement nodes $\mathbf{H}$ from $\mathbf{Z}$. Finally, BLS estimates $\mathbf{Y}$ from both $\mathbf{Z}$ and $\mathbf{H}$.

The steps to build a BLS are:
\begin{enumerate}
\item \emph{Construct the linear feature nodes\footnote{In Section~III-A of \cite{chen2018broad}, it is stated that ``\emph{In our BLS, to take the advantages of sparse autoencoder characteristics, we apply the linear inverse problem in (7) and fine-tune the initial $\mathbf{W}_{e_i}$ to obtain better features.}" However, its context and Algorithms~1-3 use randomly initialized $\mathbf{W}_{e_i}$, and do not mention exactly how the sparse autoencoder is used. Here we describe the BLS procedure according to their sample code at http://www.broadlearning.ai/, which includes the details on how the sparse autoencoder is implemented. We also compared $\mathbf{W}_{e_i}$ with and without sparse autoencoder, and found that the former indeed worked better.} $\mathbf{Z}$}. Let $n$ be the number of groups of features nodes, and $m$ be the number of features nodes in each group. We first concatenate $\mathbf{X}$ with an all-one bias vector $\mathbf{1}\in \mathbb{R}^{N\times 1}$ to form the augmented data matrix $\mathbf{X}'=[\mathbf{X}\ \mathbf{1}]$, then construct each of the $n$ groups of feature nodes, $\{\mathbf{Z}_i\}_{i=1}^n$, individually. For the $i$th group of feature nodes $\mathbf{Z}_i$, we first randomly generate uniformly distributed feature weights $\mathbf{W}_r\in \mathbb{R}^{(M+1)\times m}$ and compute the random feature nodes $\mathbf{Z}_{r_i}=\mathbf{X}'\mathbf{W}_r$, then use least absolute shrinkage and selection operator (LASSO) to obtain sparse weights $\mathbf{W}_{e_i}\in \mathbb{R}^{(M+1)\times m}$:
    \begin{align}
    \mathbf{W}_{e_i}&=\arg\min_{\mathbf{W}}(\frac12||\mathbf{Z}_{r_i}\mathbf{W}-\mathbf{X}'||_{F}^2+\lambda_1 || \mathbf{W}||_{1,1})^T, \label{eq:lasso}
    \end{align}
    where $|| \mathbf{W}||_{1,1} = \sum_{i=1}^{m}\sum_{j=1}^{M+1}\left|w_{i j}\right|$, and $\lambda_1$ is the L1 regularization coefficient. Alternating direction method of multipliers \cite{Boyd2010} is applied to solve (\ref{eq:lasso}). Then, we construct $\mathbf{Z}_i=\mathbf{X}'\mathbf{W}_{e_i}$, and $\mathbf{Z} = [\mathbf{Z}_1,...,\mathbf{Z}_n]$.

\item \emph{Construct the $k$ nonlinear enhancement nodes $\mathbf{H}$}. Let $\xi$ be the hyperbolic tangent sigmoid function, i.e.,
    \begin{align}
    \xi(x)=\frac{2}{1+e^{-2x}}-1 \label{eq:xi}
    \end{align}
    then
\begin{align}
\mathbf{H}'&=[\mathbf{Z},\mathbf{1}]\mathbf{W}_h, \label{eq:h'}\\
\mathbf{H}&=\xi\left(\frac{s\mathbf{H}'}{\max(\operatorname{abs}(\mathbf{H}'))}\right), \label{eq:h}
\end{align}
where $\mathbf{W}_h\in \mathbb{R}^{(nm+1)\times k}$ is a matrix of the orthonormal bases of a randomly generated uniformly distributed weight matrix in $\mathbb{R}^{(nm+1)\times k}$, $s$ is a scalar normalization factor, and $\max(\operatorname{abs}(\mathbf{H}'))$ is the maximum absolute value of all elements in $\mathbf{H}'$. The goal of $\frac{s\mathbf{H}'}{\max(\operatorname{abs}(\mathbf{H}'))}$ is to constrain the input to $\xi$ to $[-s,s]$, i.e., it performs normalization.

\item \emph{Calculate $\mathbf{W}_o\in \mathbb{R}^{(nm+k)\times C}$, the weights from $[\mathbf{Z}, \mathbf{H}]$ to $\mathbf{Y}$}. Ridge regression is used to compute $\mathbf{W}_o$, i.e.,
\begin{align}
\mathbf{W}_o&=\arg\min_{\mathbf{W}}(||[\mathbf{Z},\mathbf{H}]\mathbf{W}-\mathbf{Y}
||_{F}^2+\lambda_2||\mathbf{W}||_{F}^2)\nonumber \\
&=(\lambda_2 \mathbf{I}+ [\mathbf{Z},\mathbf{H}]^T[\mathbf{Z},\mathbf{H}])^{-1}[\mathbf{Z},\mathbf{H}]^T\mathbf{Y}, \label{eq:WoBLS}
\end{align}
where $\lambda_2$ is the L2 regularization coefficient.
\end{enumerate}

The pseudocode of BLS is given in Algorithm~\ref{Alg:BLS}. Through $n$ random feature weight matrices $\mathbf{W}_r$ and L1 regularization, BLS extracts multiple sets of diverse linear de-noised features $\mathbf{Z}$ (which help increase its generalization ability). Then, orthogonal mapping and sigmoid functions are used to construct the enhancement nodes $\mathbf{H}$ to introduce more nonlinearity (which help increase its model fitting power). Finally, $\mathbf{Z}$ and $\mathbf{H}$ are concatenated as the features for predicting $\mathbf{Y}$.

\begin{algorithm}
\caption{The single-view BLS training algorithm \cite{chen2018broad}.}\label{Alg:BLS}
\begin{algorithmic}
\REQUIRE $\mathbf{X}\in \mathbb{R}^{N\times M}$, the training data matrix; \\
\hspace*{8mm} $\mathbf{Y}\in \mathbb{R}^{N\times C}$, the corresponding one-hot coding label matrix of $\mathbf{X}$;\\
\hspace*{8mm} $n$, the number of feature node groups;\\
\hspace*{8mm} $m$, the number of feature nodes in each group;\\
\hspace*{8mm} $k$, the number of enhancement nodes;\\
\hspace*{8mm} $s$, the normalization factor;\\
\hspace*{8mm} $\lambda_1$, the L1 regularization coefficient for determining $\mathbf{W}_{e_i}$;\\
\hspace*{8mm} $\lambda_2$, the L2 regularization coefficient for determining $\mathbf{W}_o$.
\ENSURE BLS weight matrices $\mathbf{W}_{e_i}\in \mathbb{R}^{(M+1)\times m}$ ($i=1,...,n$), $\mathbf{W}_h\in \mathbb{R}^{(nm+1)\times k}$, and $\mathbf{W}_o\in \mathbb{R}^{(nm+k)\times C}$.
\STATE Construct $\mathbf{X}'=[\mathbf{X}, \mathbf{1}]$
\FOR{$i=1$ to $n$}
\STATE Initialize $\mathbf{W}_r$ randomly;
\STATE Calculate $\mathbf{Z}_{r_i}=\mathbf{X}'\mathbf{W}_r$;
\STATE Calculate $\mathbf{W}_{e_i}$ using (\ref{eq:lasso});
\STATE Calculate feature nodes $\mathbf{Z}_i=\mathbf{X}'\mathbf{W}_{e_i}$;
\ENDFOR
\STATE Construct $\mathbf{Z} = [\mathbf{Z}_1,...,\mathbf{Z}_n]$;
\STATE Construct an orthonormal basis matrix $\mathbf{W}_h\in \mathbb{R}^{(nm+1)\times k}$ from a randomly generated matrix in $\mathbb{R}^{(nm+1)\times k}$;
\STATE Calculate the enhancement nodes $\mathbf{H}$ using (\ref{eq:h'}) and (\ref{eq:h});
\STATE Calculate $\mathbf{W}_o$ using (\ref{eq:WoBLS}).
\end{algorithmic}
\end{algorithm}

\subsection{Multi-View Broad Learning System (MvBLS)} \label{sect:MvBLS}

BLS has achieved comparable performance, with much less computational cost, with deep learning approaches on two single-view image datasets \cite{chen2018broad}. However, it is not optimized for multi-view data. This subsection extends single-view BLS to multi-view.

\begin{figure}[htpb] \centering
\includegraphics[width=.95\linewidth,clip]{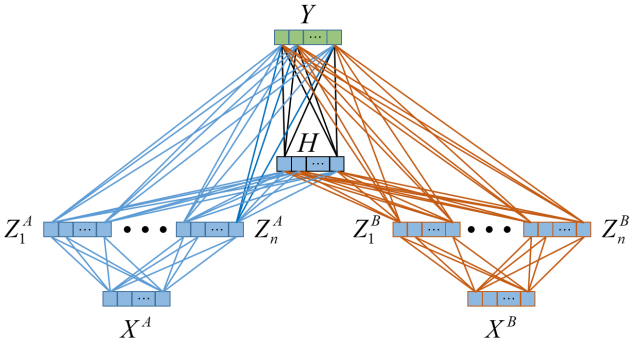}
\caption{Architecture of the proposed MvBLS. Diverse linear de-noised features $\mathbf{Z}^A$ and $\mathbf{Z}^B$ are extracted from Views~A and B, respectively. $\mathbf{Z}^A$ and $\mathbf{Z}^B$ are then mapped into nonlinear features $\mathbf{H}$. $\mathbf{Z}^A$, $\mathbf{Z}^B$ and $\mathbf{H}$ are next concatenated to predict $\mathbf{Y}$.} \label{fig:MvBLS1}
\end{figure}

The architecture of the proposed MvBLS is shown in Fig.~\ref{fig:MvBLS1}. Without loss of generality, we only consider two views. The extension to more than two views is straightforward. The general idea is to construct the linear de-noised feature nodes of each view separately, concatenate the feature nodes from all views to construct the nonlinear enhancement nodes, and finally fuse the feature nodes and enhancement nodes together for prediction. By separating the two views in the first layer of the MvBLS and optimizing $\mathbf{Z}^A$ and $\mathbf{Z}^B$ separately, we may obtain better features than optimizing $\mathbf{Z}=[\mathbf{Z}^A\ \mathbf{Z}^B]$ directly (as in the case that we concatenate $\mathbf{X}^A$ and $\mathbf{X}^B$ and feed them altogether into a single BLS), because $\mathbf{Z}$ may be too long to be optimized effectively.

Let the two views be $A$ and $B$, the corresponding data matrices be $\mathbf{X}^A\in \mathbb{R}^{N\times M_A}$ and $\mathbf{X}^B\in \mathbb{R}^{N\times M_B}$ ($M_A$ and $M_B$ are the feature dimensionality of Views $A$ and $B$, respectively), and the shared label matrix be $\mathbf{Y}\in \mathbb{R}^{N\times C}$. The procedure for constructing the MvBLS is:
\begin{enumerate}
\item \emph{Construct the feature nodes $\mathbf{Z}^A=[\mathbf{Z}_1^A,...,\mathbf{Z}_n^A]$ for View~$A$, and $\mathbf{Z}^B=[\mathbf{Z}_1^B,...,\mathbf{Z}_n^B]$ for View~$B$, using Step~(1) of Algorithm~1}.

\item \emph{Construct the enhancement nodes $\mathbf{H}$, using the concatenated feature nodes $[\mathbf{Z}^A, \mathbf{Z}^B]$ from both views and Step~(2) of Algorithm~1}.

\item \emph{Calculate $\mathbf{W}_o\in \mathbb{R}^{(2nm+k)\times C}$, the weights from $[\mathbf{Z}^A, \mathbf{Z}^B, \mathbf{H}]$ to $\mathbf{Y}$}. Again, ridge regression is used to compute $\mathbf{W}_o$. Let $\mathbf{Z}'=[\mathbf{Z}^A, \mathbf{Z}^B, \mathbf{H}]$. Then,
\begin{align}
\mathbf{W}_o&=\arg\min_{\mathbf{W}}(||\mathbf{Z}'\mathbf{W}-\mathbf{Y}||_{F}^2+\lambda_2||\mathbf{W}||_{F}^2)
\nonumber \\
&=(\lambda_2 \mathbf{I}+ \mathbf{Z}'^T\mathbf{Z}')^{-1}\mathbf{Z}'^T\mathbf{Y}, \label{eq:WoMVBLS}
\end{align}
where $\lambda_2$ is the L2 regularization coefficient.
\end{enumerate}

The pseudocode for MvBLS is shown in Algorithm~\ref{alg:MvBLS1}.

\begin{algorithm}
\caption{The proposed MvBLS for two views.} \label{alg:MvBLS1}
\begin{algorithmic}
\REQUIRE $\mathbf{X}^A\in \mathbb{R}^{N\times M_A}$, the training data matrix for View~$A$; \\
\hspace*{8mm} $\mathbf{X}^B\in \mathbb{R}^{N\times M_B}$, the training data matrix for View~$B$; \\
\hspace*{8mm} $\mathbf{Y}\in \mathbb{R}^{N\times C}$, the corresponding one-hot coding label matrix;\\
\hspace*{8mm} $n$, the number of feature node groups;\\
\hspace*{8mm} $m$, the number of feature nodes in each group;\\
\hspace*{8mm} $k$, the number of enhancement nodes;\\
\hspace*{8mm} $s$, the normalization factor;\\
\hspace*{8mm} $\lambda_1$, the L1 regularization coefficient for determining $\mathbf{W}_{e_i}$;\\
\hspace*{8mm} $\lambda_2$, the L2 regularization coefficient for determining $\mathbf{W}_o$.
\ENSURE MvBLS weight matrices $\mathbf{W}_{e_i}^A\in \mathbb{R}^{(M_A+1)\times m}$, $\mathbf{W}_{e_i}^B\in \mathbb{R}^{(M_B+1)\times m}$ ($i=1,...,n$), $\mathbf{W}_h\in \mathbb{R}^{(2nm+1)\times k}$, and $\mathbf{W}_o\in \mathbb{R}^{(2nm+k)\times C}$
\STATE Calculate $\mathbf{W}_{e_i}^A$ and $\mathbf{W}_{e_i}^B$ using $\mathbf{X}^A$, $\mathbf{X}^B$, $\mathbf{Y}$, and Step~(1) of BLS to construct the feature nodes $\mathbf{Z}^A$ and $\mathbf{Z}^B$;
\STATE Calculate $\mathbf{W}_h$ using $\mathbf{Z}^A$, $\mathbf{Z}^B$, and Step~(2) of BLS;
\STATE Calculate $\mathbf{W}_o$ using (\ref{eq:WoMVBLS}).
\end{algorithmic}
\end{algorithm}

\section{Experiment and Results}

This section applies BLS and MvBLS to monkey oculomotor decision classification, and compares their performance with those using several classical and state-of-the-art single-view and multi-view learning approaches.

\subsection{The Neurophysiology Experiment}

The invasive neurophysiological experimental setup used here and animal behavior were reported in \cite{Chen2015Sequential}. All animal care and experimental procedures were in compliance with the US Public Health Service policy on the humane care and use of laboratory animals, and were approved by Johns Hopkins University Animal Care and Use Committee.

\begin{figure}[htpb] \centering
\includegraphics[width=\linewidth,clip]{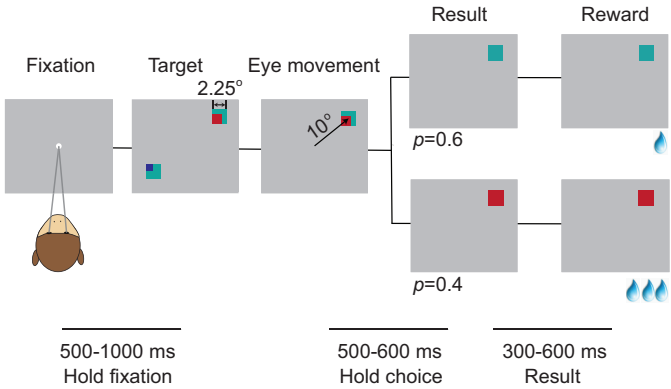}
\caption{Sequence of events in the oculomotor gambling task. The figure is modified from Figure~2B in  \cite{Chen2015Sequential}. In the `target' step, two visual cues appeared in two random directions. After the visual cues were presented, the monkey made a choice between the two options by making a saccade to the corresponding visual cue (`saccade' step), indicted by the black arrow (the black arrow was artificially added to the figure to better explain the experiment design, but it was not included in the actual visual cue displayed to the monkeys). The lines at the bottom indicate the duration of various time periods in the gambling task.} \label{fig:trial}
\end{figure}

\begin{figure}[htpb] \centering
\includegraphics[width=.6\linewidth,clip]{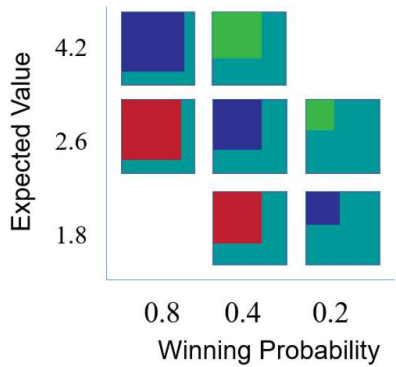}
\caption{The seven visual cues used in the gambling task. The figure is modified from Figure~2A in \cite{Chen2015Sequential}. Four different colors (cyan, red, blue, and green) indicated different amounts of reward (increasing from 1, 3, 5 to 9 units of water, where 1 unit equaled 30 $\mu$L). For example, the expected value of the right green/cyan target is: 9 units (reward amount) $\times$ 0.2 (reward probability) $+$ 1 unit (reward amount) $\times$ 0.8 (reward probability) $=$ 2.6 units. } \label{fig:VisualCues}
\end{figure}

Two male rhesus monkeys (Monkey A: 7.5 kg; Monkey I: 7.2 kg) were trained to perform an oculomotor gambling task, as shown in Fig.~\ref{fig:trial}. In each trial, the monkeys chose between two gamble options by making an eye movement (saccade) towards one of the visual cues. Two visual cues were randomly presented in two of four fixed locations (top right, bottom right, top left, and bottom left). Each cue was comprised of two colors (from a four-color library of cyan, red, blue, and green) and each color was associated with an amount of reward (1, 3, 5 to 9 units of water respectively, where 1 unit equaled 30 $\mu$L). The background color of a visual cue was cyan (small reward), and the foreground color was either red, or blue, or green (larger reward). The proportion of the two colors represented the probability of winning the corresponding reward. For the red/cyan target in Fig.~\ref{fig:trial}, there was a 60\% probability of having one unit of water (cyan color), and a 40\% probability of having three units of water (red color). The expected reward value would then be $0.6\times 1 + 0.4\times 3=1.8$ units of water. There were a total of seven gamble options, representing three different expected reward values, as shown in Fig.~\ref{fig:VisualCues}.

Neural signals from the monkeys' supplementary eye field in the medial frontal cortex were recorded with one or more tungsten electrodes, and the monkeys' corresponding choices were recorded using an eye tracking system (Eye Link, SR Research Ltd, Ottawa, Canada).

The goal of our study was to investigate the task of decoding eye movements (choice intention) from neural signals in the primate medial frontal cortex, which are causally involved in risky decisions \cite{Chen2018a}.

\subsection{Datasets}

Forty-five datasets \cite{Chen2015Sequential} were recorded from 45 experiment sessions from the two monkeys (33 from Monkey A, and 12 from Monkey I). Their statistics are shown in Table~\ref{tab:data}, where $d_1$-$d_4$ denote the four different saccade directions (classes), $nE$ the number of electrodes in recording the LFPs, and $nU$ the number of units in recording the spikes.

For each recording, electrodes were lowered into the monkeys' supplementary eye field using electric microdrives. While the monkeys were preforming the task, activity was recorded extracellularly using 1 to 4 tungsten microelectrodes with an impedance of 2-4 M$\Omega$s (Frederick Haer, Bowdoinham, ME, USA) spaced 1-3 mm apart. Neural activity was measured against a local reference, a stainless steel guide tube, which carried the electrode array and was positioned above the dura.

At the preamplifier stage, signals were processed with 0.5 Hz 1-pole high-pass and 8k Hz 4-pole low-pass anti-aliasing Bessel filters, and then divided into two streams for the recording of LFPs and spiking activity. The stream used for LFP recording was amplified (500-2000 gain), processed by a 4-pole 200 Hz low-pass Bessel filter, and sampled at 1000 Hz. The stream used for spike detection was processed by a 4-pole Bessel high-pass filter (300 Hz), a 2-pole Bessel low-passed filter (6000 Hz), and was sampled at 40k Hz. Up to four template spikes were identified using principal component analysis. The spiking activity was subsequently analyzed off-line. To maximize task related information in the spiking activity, all spike waveforms above the sorting threshold were used in the analysis. They included spiking activities from single units, multi-units, unsorted units, and units with very low firing rates.

\begin{figure}[htpb] \centering
\subfigure[]{\includegraphics[width=.38\linewidth,clip]{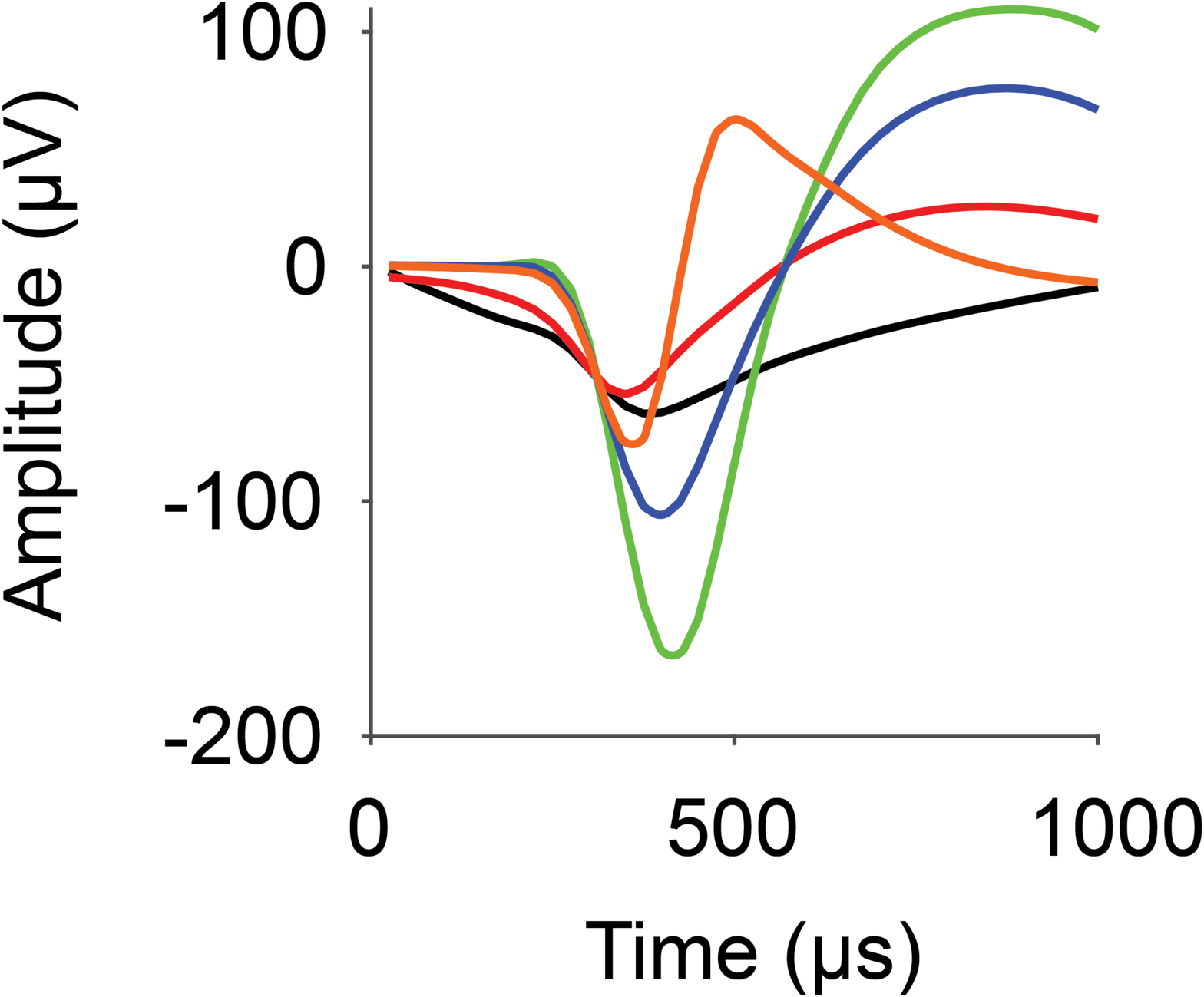}}
\subfigure[]{\includegraphics[width=.57\linewidth,clip]{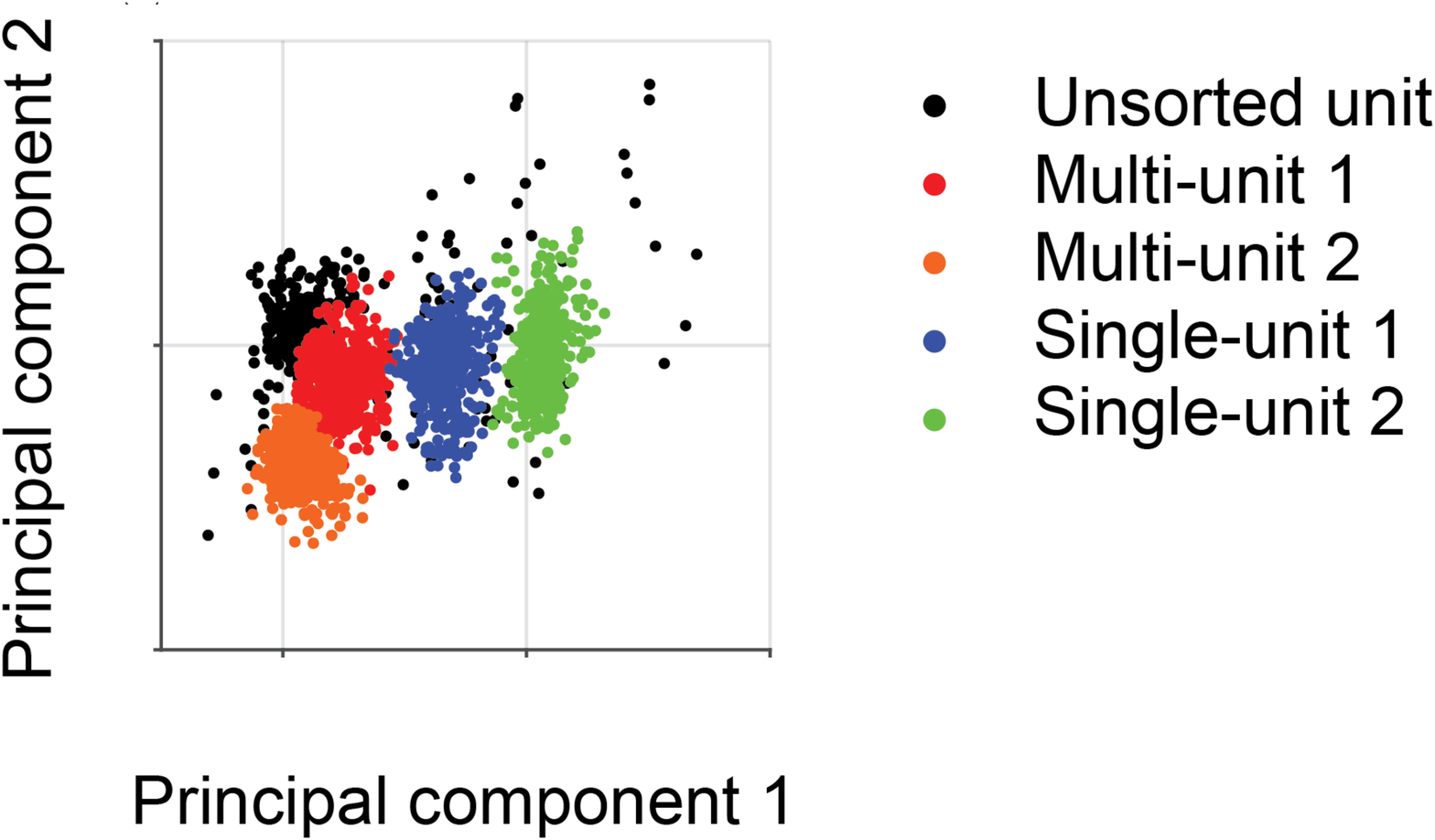}}
\subfigure[]{\includegraphics[width=.3\linewidth,clip]{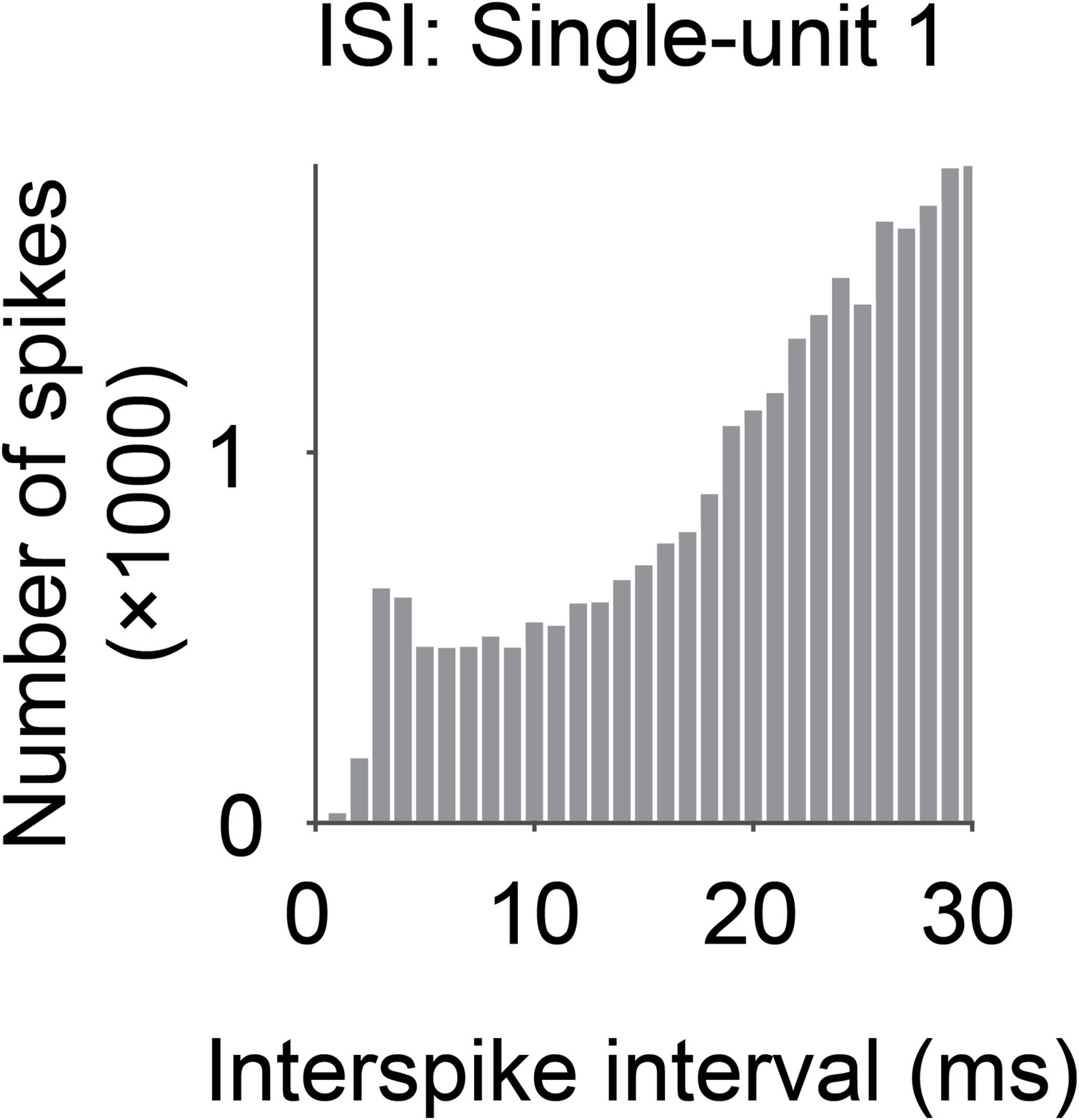}}
\subfigure[]{\includegraphics[width=.25\linewidth,clip]{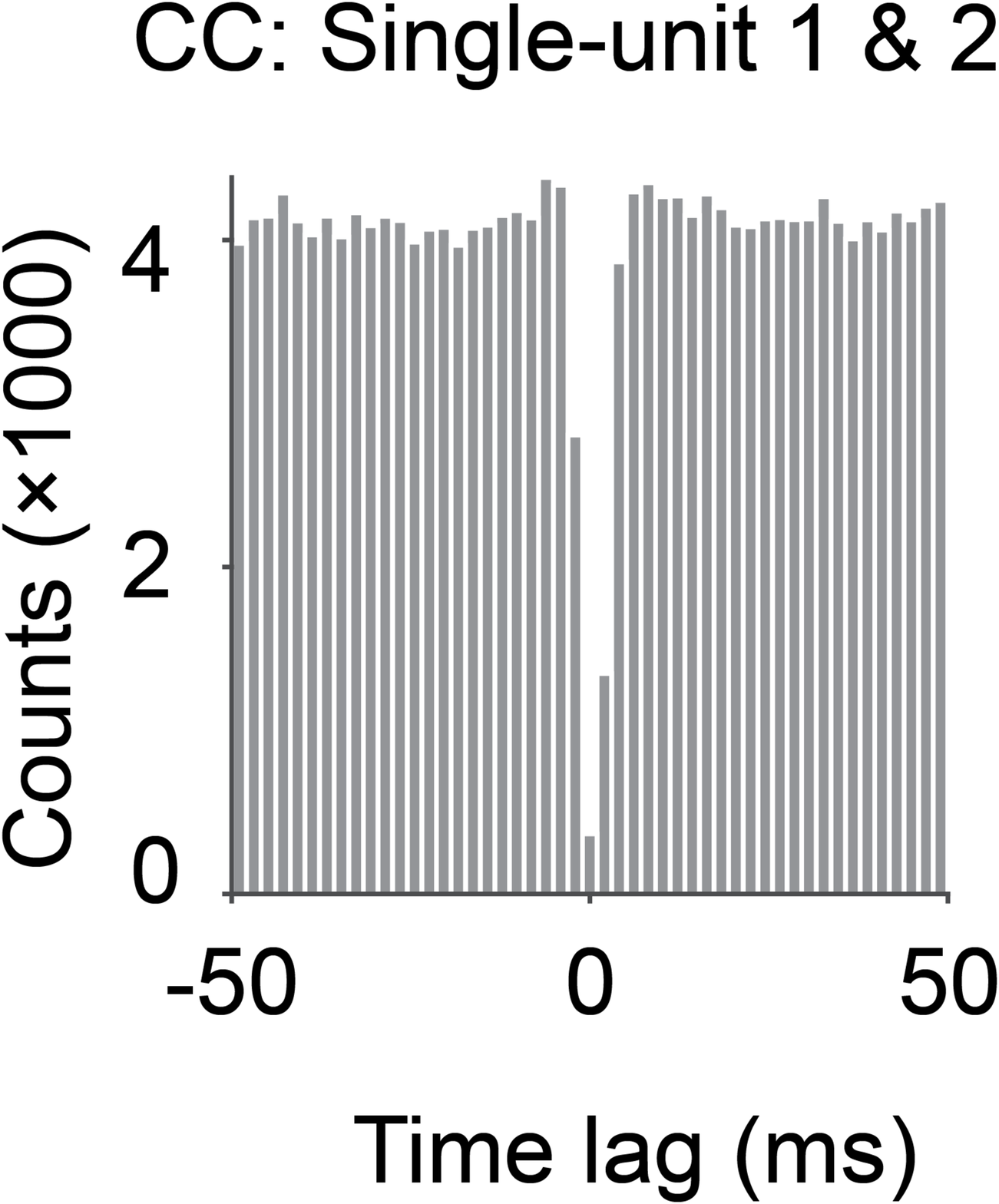}}
\subfigure[]{\includegraphics[width=.42\linewidth,clip]{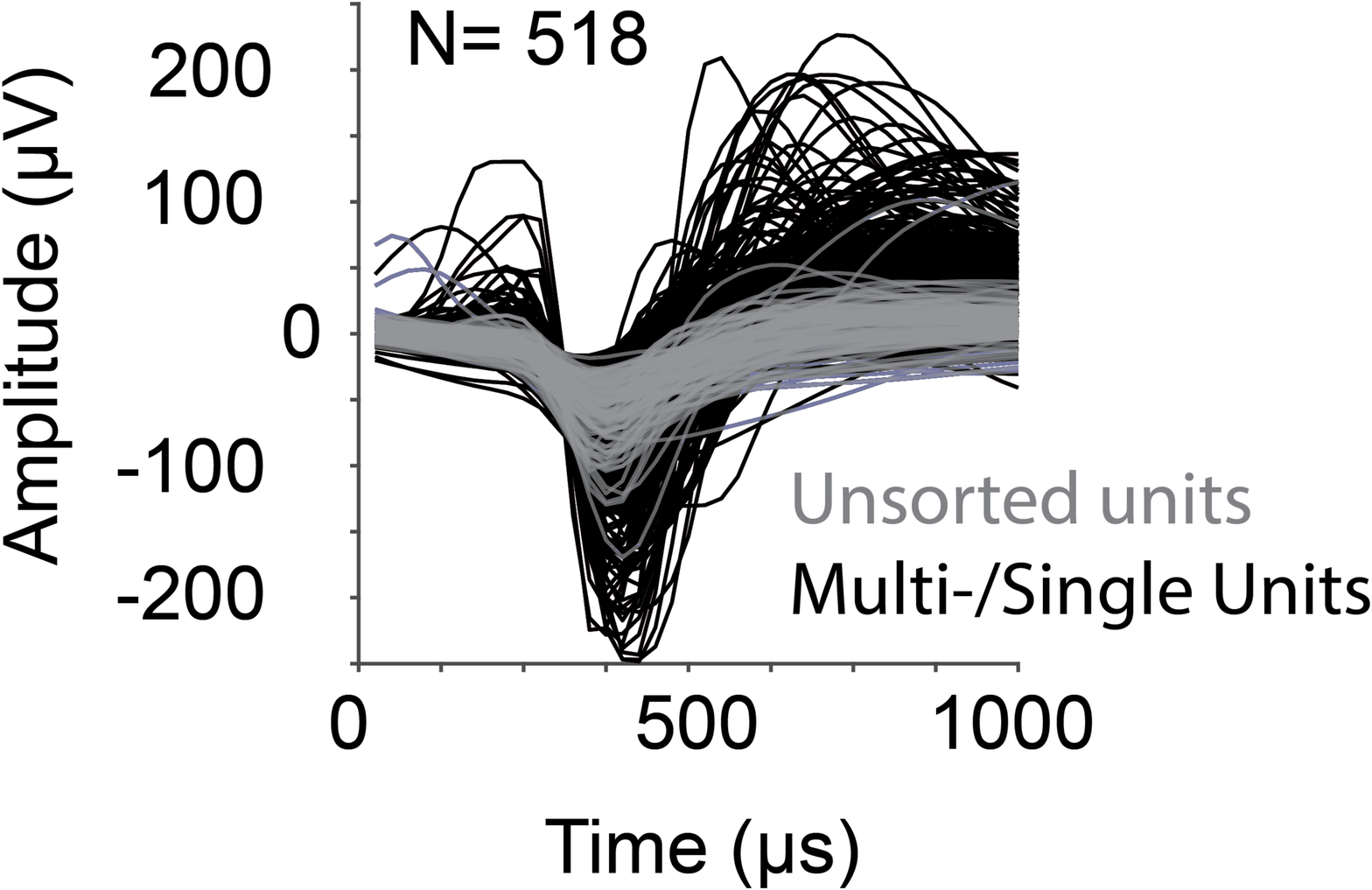}}
\caption{A representative recording (a-d) and population spike waveforms from all experimental sessions. (a) Average spike waveforms for unsorted unit, multi-units, and single-units recorded from the same electrode; (b) scatter plot of spike waveforms using the first two principal components; (c) ISI histogram of a single unit (bin width 1 ms); (d) cross-correlograms between two single-units; and (e) average spike waveforms of all recorded units (unsorted units are in grey).} \label{fig:units}
\end{figure}

Fig.~\ref{fig:units} illustrates two single units, two multi-units and an unsorted units we identified in a representative recording. A single-unit is defined as a well-isolated unit (see Fig.~\ref{fig:units}(b)) whose interspike interval (ISI) violation rate (see Fig.~\ref{fig:units}(c)) is smaller than 0.02. A multi-unit is defined as a unit with a clear template waveform (online or offline); however, it is not well isolated from other clusters of waveforms (see Fig.~\ref{fig:units}(b)) and has a high ISI violation rate ($\ge0.02$). An unsorted unit is defined by the Plexon Offline Sorter, whose spike waveforms do not belong to any category of sorted waveforms (see Figs.~\ref{fig:units}(b) and (e)) \cite{Chung2017,Kloosterman2014,Trainito2019}. Finally, spikes were counted within each millisecond bin. Therefore, the temporal frequency of the spiking activity was 1000 Hz.

In the remainder of the paper, we refer both multi-units and unsorted units as multi-units for simplicity.

%\begin{figure}[htpb] \centering
%\includegraphics[width=\linewidth,clip]{FigureX.eps}
%\caption{Representative recordings and population spike-waveforms. (a) Average spike waveforms for unsorted unit, multi-units, and single-units recorded from the same electrode; (b) scatter plot of spike waveforms using the first two principal components; (c) ISI histogram of a single unit (bin width 1 ms); (d) cross-correlograms between two single-units; and (e) average spike waveforms of all recorded units (unsorted units are in grey).} \label{fig:units}
%\end{figure}

\begin{table}[htpb] \centering \setlength{\tabcolsep}{1.5mm}
\caption{Statistics of the 45 datasets.} \label{tab:data}
\begin{tabular}{c|cccccr|c|r|r|r}
\hline
\multirow{2}{*}{Dataset} & \multicolumn{6}{c|}{Number of trials}       & \multirow{2}{*}{$nE$} & \multirow{2}{*}{$nU$} & \multirow{2}{*}{$nS$}  & \multirow{2}{*}{$nM$}  \\ \cline{2-7}
                         & $d_1$ & $d_2$ & $d_3$ & $d_4$ & Avg & Std   &                       &                       &                        &                        \\ \hline
1                        & 218   & 236   & 169   & 142   & 191 & 43.35 & 4                     & 20                    & 5                      & 15                     \\
2                        & 255   & 259   & 195   & 189   & 225 & 37.64 & 3                     & 16                    & 5                      & 11                     \\
3                        & 275   & 310   & 256   & 195   & 259 & 48.17 & 3                     & 14                    & 5                      & 8                      \\
4                        & 198   & 187   & 197   & 141   & 181 & 26.96 & 3                     & 13                    & 7                      & 6                      \\
5                        & 203   & 210   & 146   & 153   & 178 & 33.16 & 3                     & 13                    & 6                      & 7                      \\
6                        & 218   & 231   & 213   & 190   & 213 & 17.11 & 3                     & 15                    & 10                     & 5                      \\
7                        & 159   & 184   & 161   & 142   & 162 & 17.25 & 3                     & 16                    & 10                     & 6                      \\
8                        & 170   & 193   & 188   & 164   & 179 & 13.94 & 3                     & 14                    & 9                      & 5                      \\
9                        & 194   & 183   & 197   & 177   & 188 & 9.36  & 3                     & 11                    & 4                      & 7                      \\
10                       & 222   & 249   & 220   & 195   & 222 & 22.07 & 3                     & 16                    & 7                      & 9                      \\
11                       & 224   & 235   & 242   & 212   & 228 & 13.12 & 3                     & 14                    & 7                      & 7                      \\
12                       & 121   & 114   & 140   & 129   & 126 & 11.17 & 2                     & 10                    & 4                      & 6                      \\
13                       & 193   & 177   & 178   & 189   & 184 & 7.97  & 3                     & 14                    & 5                      & 9                      \\
14                       & 251   & 220   & 201   & 192   & 216 & 26.09 & 2                     & 10                    & 2                      & 8                      \\
15                       & 227   & 211   & 184   & 176   & 200 & 23.67 & 2                     & 7                     & 1                      & 6                      \\
16                       & 207   & 188   & 183   & 156   & 184 & 21.05 & 2                     & 8                     & 2                      & 6                      \\
17                       & 225   & 203   & 131   & 173   & 183 & 40.69 & 2                     & 8                     & 0                      & 8                      \\
18                       & 228   & 208   & 194   & 188   & 205 & 17.77 & 2                     & 10                    & 1                      & 9                      \\
19                       & 185   & 166   & 149   & 148   & 162 & 17.42 & 3                     & 13                    & 3                      & 10                     \\
20                       & 169   & 147   & 145   & 164   & 156 & 12.04 & 1                     & 3                     & 1                      & 2                      \\
21                       & 170   & 151   & 117   & 137   & 144 & 22.38 & 1                     & 4                     & 1                      & 3                      \\
22                       & 163   & 144   & 102   & 126   & 134 & 26.00 & 1                     & 5                     & 1                      & 4                      \\
23                       & 193   & 192   & 171   & 182   & 185 & 10.28 & 1                     & 4                     & 3                      & 1                      \\
24                       & 196   & 183   & 164   & 172   & 179 & 13.89 & 1                     & 3                     & 1                      & 2                      \\
25                       & 148   & 138   & 104   & 138   & 132 & 19.25 & 2                     & 7                     & 3                      & 4                      \\
26                       & 209   & 166   & 129   & 182   & 172 & 33.43 & 2                     & 9                     & 5                      & 4                      \\
27                       & 218   & 168   & 133   & 210   & 182 & 39.48 & 2                     & 9                     & 1                      & 8                      \\
28                       & 183   & 161   & 118   & 164   & 157 & 27.45 & 1                     & 2                     & 0                      & 2                      \\
29                       & 198   & 173   & 111   & 170   & 163 & 36.87 & 3                     & 13                    & 8                      & 5                      \\
30                       & 188   & 181   & 144   & 143   & 164 & 23.85 & 3                     & 12                    & 4                      & 8                      \\
31                       & 216   & 189   & 174   & 212   & 198 & 19.81 & 3                     & 14                    & 6                      & 8                      \\
32                       & 213   & 206   & 137   & 213   & 192 & 36.98 & 3                     & 14                    & 9                      & 5                      \\
33                       & 198   & 174   & 116   & 179   & 167 & 35.38 & 3                     & 13                    & 5                      & 7                      \\
34                       & 212   & 188   & 115   & 178   & 173 & 41.37 & 3                     & 14                    & 7                      & 7                      \\
35                       & 304   & 274   & 168   & 274   & 255 & 59.70 & 3                     & 14                    & 3                      & 11                     \\
36                       & 274   & 222   & 134   & 245   & 219 & 60.37 & 3                     & 14                    & 7                      & 7                      \\
37                       & 263   & 202   & 114   & 204   & 196 & 61.41 & 3                     & 14                    & 5                      & 9                      \\
38                       & 244   & 202   & 160   & 202   & 202 & 34.29 & 3                     & 13                    & 2                      & 11                     \\
39                       & 249   & 240   & 134   & 223   & 212 & 52.78 & 3                     & 15                    & 8                      & 7                      \\
40                       & 273   & 253   & 182   & 220   & 232 & 39.86 & 3                     & 15                    & 7                      & 8                      \\
41                       & 133   & 136   & 134   & 123   & 132 & 5.80  & 2                     & 9                     & 6                      & 3                      \\
42                       & 138   & 154   & 141   & 117   & 138 & 15.33 & 3                     & 11                    & 3                      & 8                      \\
43                       & 283   & 248   & 232   & 246   & 252 & 21.70 & 3                     & 15                    & 8                      & 7                      \\
44                       & 194   & 187   & 129   & 162   & 168 & 29.41 & 3                     & 15                    & 8                      & 7                      \\
45                       & 146   & 187   & 145   & 124   & 151 & 26.36 & 3                     & 15                    & 8                      & 7                      \\ \hline
Avg                      & 208   & 196   & 160   & 177   & 185 & 27.85 & $-$                   & $-$                   & $-$                    & $-$                    \\ \hline
\end{tabular}
\end{table}

\subsection{LFP and Spike Feature Extraction} \label{sect:LFPspikes}

In this study, single-unit and multi-unit spikes were smoothed through 100-point moving average. The LFPs and processed spikes were then epoched to $[0, 400)$ ms with 1 ms temporal resolution after target onset for each electrode/unit. The eye movement reaction times (the time between target onset and eye movement onset) for the monkeys ranged from 100 ms to 300 ms. Therefore, the monkeys finished their eye movement before the end of each trial.

Each trial had two views: the spike view and the LFP view. The spike view had $400\cdot nU$ features, where $nU$ is the number of units for spikes in Table~\ref{tab:data}. For each LFP trial from each electrode,  power spectrum density of the 400-ms signal was computed by the Welch's method with a Hamming window of 88 ms and 50\% overlap. Then, log-power in eight frequency bands (theta, 4-8 Hz; alpha, 8-12 Hz; beta 1, 12-24 Hz; beta 2, 24-34 Hz; gamma 1, 34-55 Hz; gamma 2, 65-95 Hz; gamma 3, 130-170 Hz; gamma 4, 170-200 Hz), as used in \cite{Hsieh2019}, were calculated and concatenated with the 400-ms signal as the features. Therefore, the LFP view had $408\cdot nE$ features, where $nE$ is the number of electrodes for LFPs in Table~\ref{tab:data}. Finally, both spike and LFP features were $z$-normalized.

\subsection{Algorithms} \label{sect:Algorithms}

We compared the performance of different decoding algorithms, including both single-view learning and multi-view learning approaches:

\begin{enumerate}

\item \emph{Support vector machine (SVM)}, which uses error-correcting output codes (ECOC) \cite{dietterich1994solving} for multi-class classification. SVM is a classical statistical machine learning approach, and has achieved outstanding performance in numerous applications. The box constraint $C$ was chosen from $\{10^{-6}, 10^{-4}, \cdots, 10^6\}$ by nested cross-validation. Each binary SVM classifier was solved by sequential minimal optimization (SMO) \cite{Fan2005}, and the optimization stopped if the gradient difference between upper and lower violators obtained by SMO was smaller than $0.001$.

\item \emph{Ridge classification (Ridge)}, which performs ridge regression to approximate the $\{0,1\}$ output of each class, and then classifies the input to the class with the largest output. The L2 regularization coefficient $\lambda_2$ was chosen from $\{10^{-6}, 10^{-4}, \cdots, 10^6\}$ by nested cross-validation.

\item \emph{BLS}, which has been introduced in Section~\ref{sect:BLS}. We used normalization factor $s = 0.8$ and L1 regularization coefficient $\lambda_1 = 0.001$, and selected the number of feature node groups $n$ from $\{10, 20\}$, the number of feature nodes in each group $m$ from $\{10, 20\}$, the number of enhancement nodes $k$ from $\{100, 500\}$, and the L2 regularization coefficient $\lambda_2$ from $\{10^{-6}, 10^{-4}, \cdots, 10^6\}$, using nested cross-validation. The alternating direction method of multipliers \cite{Boyd2010} used to solve (\ref{eq:lasso}) for feature nodes construction was iterative, and it terminated after 50 iterations.

\item \emph{Multi-view discriminant analysis with view-consistency (MvDA)} \cite{Kan2016}, which extends classical single-view linear discriminant analysis to multi-view learning, and adds a regularization term to enhance the view-consistency.

\item \emph{Multi-view modular discriminant analysis (MvMDA)} \cite{Cao2018}, which exploits the distance between class centers across different views.

\item \emph{MvBLS}, which has been introduced in Section~\ref{sect:MvBLS}. Its parameter tuning was the same as that for BLS.
\end{enumerate}

Note that the first three algorithms can be used for both single-view learning and multi-view learning. When they were used in multi-view learning, we simply combined the features from different views as a single view input to the classifier. The last three approaches were used in multi-view learning only. The subspace dimensionality of MvDA and MvMDA was set to three (the number of classes minus one). After subspace alignment, the subspace features of all views were concatenated and fed into an ECOC-SVM classifier, where the same nested cross-validation was conducted as in the SVM approach. Linear kernel was employed in all SVMs.

We randomly partitioned each dataset into three subsets: 60\% for training, 20\% for validation, and the remaining 20\% for test. We repeated this process 30 times on each of the 45 datasets, and recorded the test classification accuracies as our performance measure.

\subsection{4-Class Classification Using only the LFPs}

In the first experiment, we used only the LFPs in 4-class classification. The classification accuracies in the 45 sessions, each averaged over 30 cross-validation runs, are shown in the bar graph in the top panel of Fig.~\ref{fig:bar}, and also in the box plot in the top-left panel of Fig.~\ref{fig:box}. The last group of the bar plot also shows the average accuracies across the 45 sessions, whose numerical values are given in Table~\ref{tab:avg}. Each standard deviation showed in Table~\ref{tab:avg} was computed from 30 average accuracies of the 45 sessions. On average BLS slightly outperformed SVM and Ridge, which was also true in 29 and 32 out of the 45 individual sessions, respectively.

To find out whether there were statistically significant differences between different algorithms, non-parametric multiple pairwise comparison tests using Dunn's procedure \cite{Dunn1964}, with a $p$-value correction using the False Discovery Rate method \cite{Benjamini1995}, were performed on the cross-validation accuracies. The null hypothesis in each pairwise comparison was the probability of observing a randomly selected value from the first group that is larger than a randomly selected value from the second group equals $0.5$, and it was rejected if $p\ge \alpha/2$, where $\alpha=0.05$. The $p$-values, when only the LFP features were used, are shown in the first part of Table~\ref{tab:Dunn}. There was no statistically significant difference between any pair of algorithms.

In summary, we have shown that when only the LFP features were used, SVM, Ridge and BLS achieved comparable classification performances (BLS may be slightly better, but there was no statistically significant difference).

\subsection{4-Class Classification Using only the Spikes}

In the second experiment, we used only the spikes in 4-class classification. The classification accuracies in the 45 sessions, each averaged over 30 cross-validation runs, are shown in the bar graph in the middle panel of Fig.~\ref{fig:bar}, and also in the box plot in the top-right panel of Fig.~\ref{fig:box}. The last group of the bar graph shows the average accuracies across the 45 sessions, whose numerical values are also given in Table~\ref{tab:avg}. For all algorithms, using spikes only achieved better classification accuracy than using LFPs only. On average Ridge slightly outperformed BLS, which was also true in 32 individual sessions. Interestingly, the opposite held when the LFP features were used. This may indicate that LFPs and spikes encode non-identical information about the oculomotor decision.

Non-parametric multiple comparison tests were also performed, and the $p$-values are shown in the second part of Table~\ref{tab:Dunn}. There was no statistically significant difference between any pair of algorithms.

In summary, we have shown that when only the spike features were used, SVM, Ridge and BLS again achieved comparable classification performance (Ridge may be slightly better, but there was no statistically significant difference).

\subsection{4-Class Classification Using both LFPs and Spikes} \label{sect:4ClassLFPsSpikes}

\begin{sidewaysfigure} \centering
\includegraphics[width=\linewidth,clip]{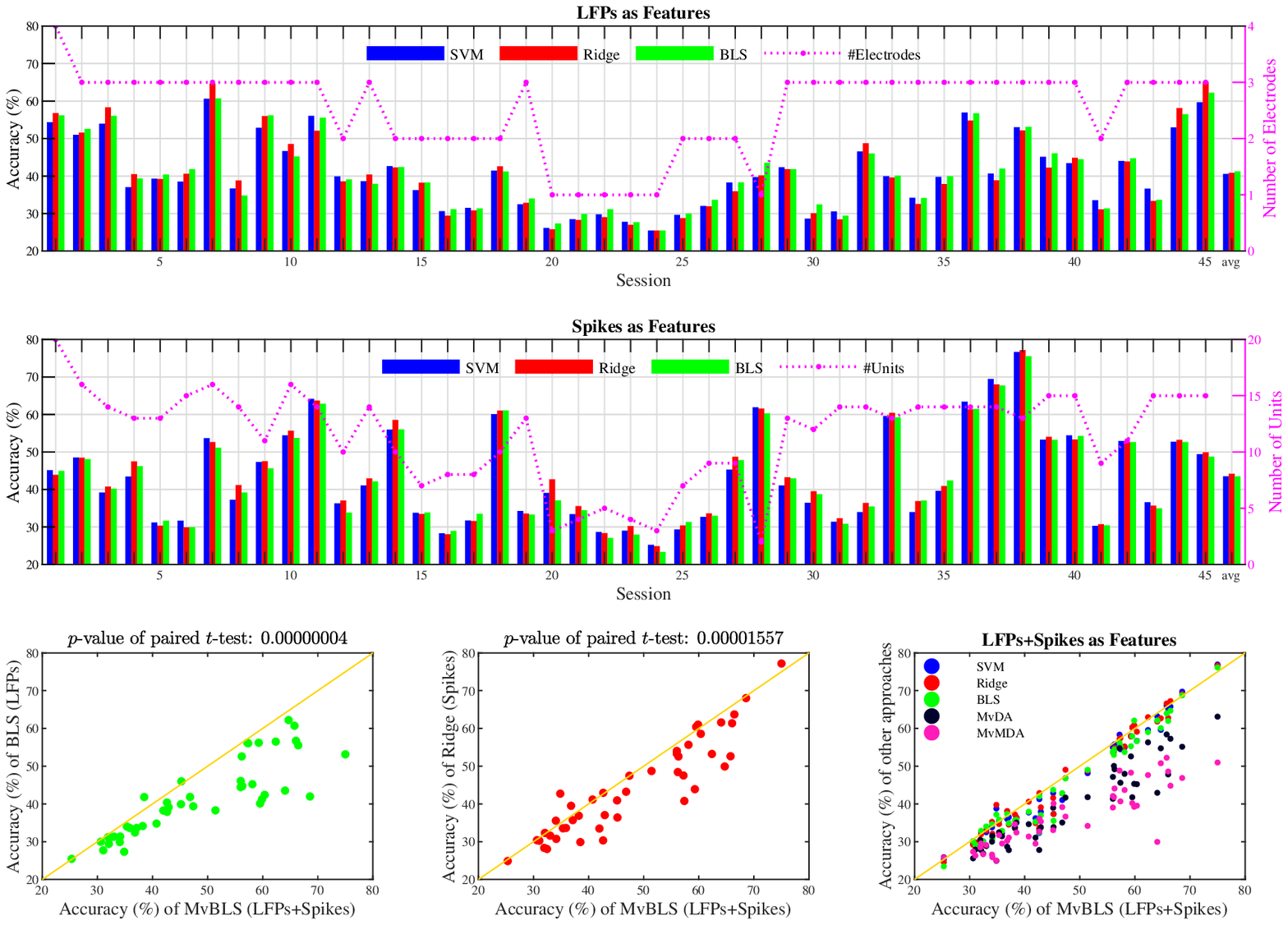}
\caption{4-class classification accuracies of different algorithms, when different features were used.} \label{fig:bar}
\end{sidewaysfigure}

\begin{figure}[htpb] \centering
\includegraphics[width=\linewidth,clip]{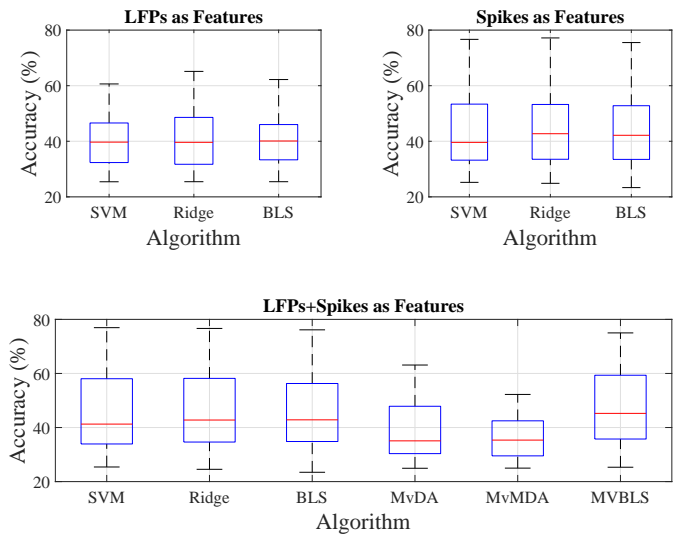}
\caption{Boxplots of the 4-class classification accuracies of different algorithms, using different features. The red line in the box indicates the median, and the bottom and top edges of the box indicate the 25th and 75th percentiles, respectively. The whiskers extend to the most extreme data points, excluding outliers.} \label{fig:box}
\end{figure}

In the third experiment, we used both LFPs and spikes in 4-class classification. The classification accuracies of MvBLS in the 45 sessions, each averaged over 30 cross-validation runs, are shown in horizontal-axis of three scatter plots in the bottom panel of Fig.~\ref{fig:bar}, and also the box plot in the bottom panel of Fig.~\ref{fig:box}. The average accuracies across the 45 sessions are given in Table~\ref{tab:avg}. Observe that:
\begin{enumerate}
\item The proposed MvBLS using LFPs+Spikes outperformed BLS using LFPs significantly, which is the best-performing approach using LFPs.
\item The proposed MvBLS using LFPs+Spikes outperformed Ridge using Spikes significantly, which is the best-performing approach using Spikes.
\item On average, the two subspace multi-view learning algorithms, i.e., MvDA and MvMDA, performed much worse than the three single-view algorithms, i.e., SVM, Ridge, and BLS.
\item On average, our proposed MvBLS outperformed the two subspace multi-view learning algorithms. This suggests that MvBLS can extract more discriminative features and better fuse them than the other two approaches.
\item On average, our proposed MvBLS also outperformed the three single-view learning algorithms. This suggests that fusing the two views in a more sophisticated way may be more advantageous than simply concatenating and feeding them into a single-view classifier.
\end{enumerate}

Non-parametric multiple comparison tests were also performed, and the $p$-values are shown in Table~\ref{tab:Dunn2}, where the statistically significant ones are marked in bold. There was statistically significant difference between MvBLS and each of the other five algorithms.

\onecolumn
\twocolumn

\begin{table*}[htpb] \centering
\caption{Mean and standard deviation of 4-class classification accuracies (\%) when different classifiers and features were used.} \label{tab:avg}
\begin{tabular}{c|cccccc}
\hline
              & SVM            & Ridge                   & BLS                     & MvDA           & MvMDA          & MvBLS                   \\ \hline
LFPs          & 40.57$\pm$0.70 & 40.85$\pm$0.70          & \textbf{41.24$\pm$0.62} & $-$            & $-$            & $-$                     \\
Spikes        & 43.49$\pm$0.62 & \textbf{44.21$\pm$0.47} & 43.48$\pm$0.59          & $-$            & $-$            & $-$                     \\
LFPs+Spikes & 46.12$\pm$0.56 & 46.51$\pm$0.41          & 45.65$\pm$0.52          & 39.49$\pm$0.50 & 36.62$\pm$0.55 & \textbf{47.94$\pm$0.62} \\ \hline
\end{tabular}
\end{table*}

\begin{table}[htpb] \centering \setlength{\tabcolsep}{4mm}
\caption{$p$-values of non-parametric multiple comparisons, when only the LFP features (first part) and the spike features (second part) were used in 4-class classification. } \label{tab:Dunn}
\begin{tabular}{l|cc|cc}
\hline
      & \multicolumn{2}{c|}{LFP Features} & \multicolumn{2}{c}{Spikes Features} \\ \cline{2-5}
      & SVM             & Ridge           & SVM               & Ridge            \\ \hline
Ridge & .4148           &                 & .1224             &                  \\
BLS   & .2607           & .1892           & .3827             & .1118            \\ \hline
\end{tabular}
\end{table}

\begin{table}[htpb] \centering \setlength{\tabcolsep}{2mm}
\caption{$p$-values of non-parametric multiple comparisons, when both LFPs and spikes were used in 4-class classification.} \label{tab:Dunn2}
\begin{tabular}{l|ccccc}
\hline
      & SVM            & Ridge          & BLS            & MvDA           & MvMDA          \\ \hline
Ridge & .2054          &                &                &                &                \\
BLS   & .2652          & .0772          &                &                &                \\
MvDA  & \textbf{.0000} & \textbf{.0000} & \textbf{.0000} &                &                \\
MvMDA & \textbf{.0000} & \textbf{.0000} & \textbf{.0000} & \textbf{.0000} &                \\
MvBLS & \textbf{.0001} & \textbf{.0019} & \textbf{.0000} & \textbf{.0000} & \textbf{.0000} \\ \hline
\end{tabular}
\end{table}

\begin{figure*}[htpb] \centering
\includegraphics[width=\linewidth,clip]{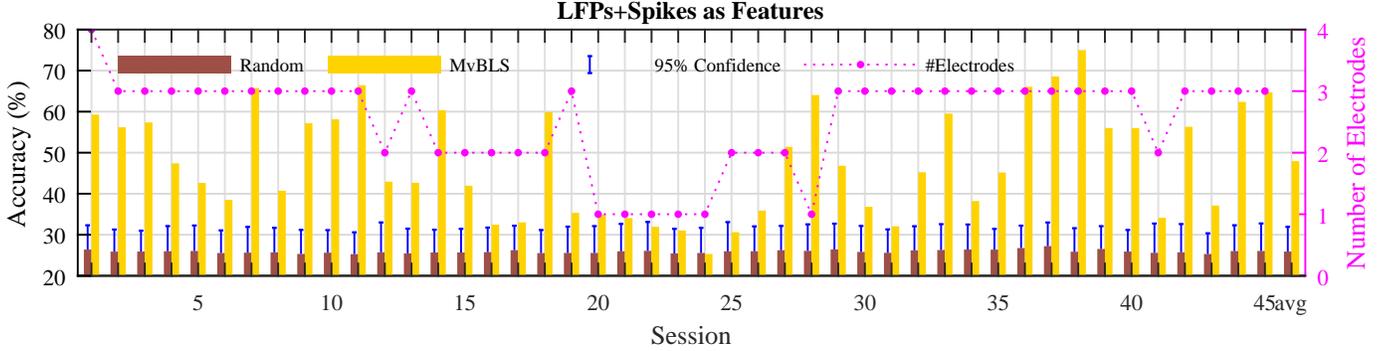}
\caption{4-class classification accuracies of permutation tests \cite{Chen2015Sequential} with 95\% confidence and MvBLS.} \label{fig:rand}
\end{figure*}

We also compared MvBLS with random guess in 4-class classification. The results are shown in Fig.~\ref{fig:rand}. Note that the random guess approach obtained slightly different accuracies in different sessions (not always exactly 25\% in 4-class classification), because different classes had different numbers of trials. On average the 4-class classification accuracy of MvBLS was about twice of that of random guess (47.94\% vs 25.04\%), suggesting that a sophisticated machine learning approach like MvBLS can indeed mine useful information from LFPs and spikes.

Finally, to validate if LFPs and spikes do contain complementary information in from another perspective, we counted the number of sessions that LFPs+Spikes achieved better performance than LFPs or spikes only, and show the results in Table~\ref{tab:sessions}. Regardless of which classifier was used, LFPs+Spikes always outperformed LFPs or Spikes alone, in most sessions. Moreover, when MvBLS was used, LFPs+Spikes outperformed the best LFPs performance (among SVM, Ridge and BLS) in 40 sessions (88.89\%), and the best Spikes performance in 32 sessions (71.11\%).

\begin{table}[htpb] \centering \setlength{\tabcolsep}{2mm}
\caption{Number and percentage of sessions (among the 45 sessions) that LFPs+Spikes outperformed a single modality of feature alone.} \label{tab:sessions}
\begin{tabular}{l|cc}
\hline
      & \multicolumn{2}{|c}{Number of sessions LFP+Spikes outperformed}\\ \cline{2-3}
      & LFPs            & Spikes                  \\ \hline
SVM   & 31 (68.89\%)    & 40 (88.89\%) \\
Ridge & 37 (82.22\%)    & 28 (62.22\%) \\
BLS   & 31 (68.89\%)    & 35 (77.78\%) \\ \hline
\end{tabular}
\end{table}

In summary, we have shown that LFPs and spikes contain complementary information about the brain's oculomotor decision, and our proposed MvBLS can better fuse these features than several classical and state-of-the-art single-view and multi-view learning approaches.

\section{Discussions}

This section presents some additional discussions on the proposed MvBLS.

\subsection{4-Class Classification Accuracy versus the Number of Electrodes/Units}

Figs.~\ref{fig:bar} and \ref{fig:rand} show that sometimes the 4-class classification accuracy was very low, e.g., close to random guess (25.91\%) or permutation tests \cite{Chen2015Sequential} with 95\% confidence (31.93\%). The main reason is that the number of electrodes/units was small in these cases. For example, the top panel of Fig.~\ref{fig:bar} also shows the number of LFP electrodes in different sessions. It has a strong correlation with the classification accuracy, regardless of which classification algorithm was used. Particularly, the sessions with the lowest classification accuracy (Sessions~20-24) had the smallest number of electrodes. The middle panel of Fig.~\ref{fig:bar} shows the number of units in different sessions. Similar patterns can be observed.

Next, we performed a deeper investigation on how the number of electrodes in LFPs and the number of units in spikes affected the performance of BLS\footnote{We studied LFPs and spikes separately. Each time there was only one view, and hence MvBLS degraded to BLS.}.

\begin{figure}[htpb] \centering
\subfigure[]{\label{fig:ele}     \includegraphics[width=.8\linewidth,clip]{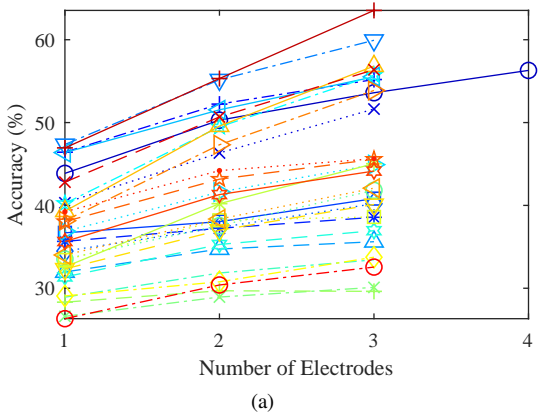}}
\subfigure[]{\label{fig:unit}     \includegraphics[width=.8\linewidth,clip]{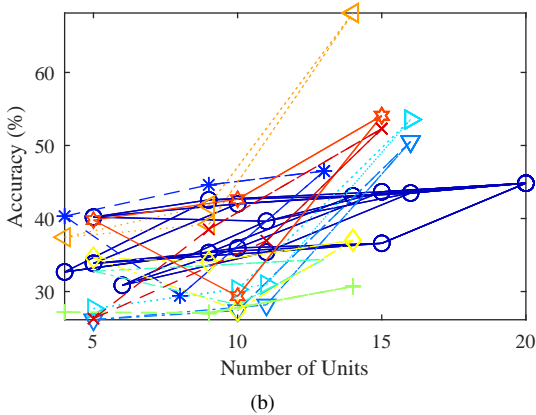}}
\caption{BLS classification (4-class) accuracy versus (a) the number of electrodes, and (b) the number of units. In (a), each curve represents a different dataset. In (b), curves with the same end-point are from the same dataset.} \label{fig:BLSvsEleUnit}
\end{figure}

The LFPs were studied first. We identified all datasets with three or more electrodes, and considered each one separately. A dataset with three electrodes is used as an example to illustrate our experimental procedure. In the experiment, we randomly partitioned the dataset into 60\% training, 20\% validation, and 20\% test. Then, we increased the number of chosen electrodes $k$ from one to two and then to three; for each $k$, we used LFPs from the corresponding electrodes to trained a BLS and compute its test accuracy. All possible combinations of $k$ electrodes were considered, and the average test accuracy was computed. After that, we repeated the data partition 30 times and computed the grand average test accuracies, as shown in Fig.~\ref{fig:ele}. Intuitively, the classification accuracy increased with the number of electrodes. We would expect that higher classification accuracy could be obtained with more electrodes.

We next studied the decoding performance with spikes. The results are shown in Fig.~\ref{fig:unit}. The experimental procedure was very similar to that for the LFPs, except one difference: the number of units associated with different electrodes were generally different, so the total number of units from $k$ electrodes could have different values. Take the first dataset with four electrodes (they had 4, 5, 5 and 6 units, respectively) as an example. All possible combinations of the 4 electrodes were considered. There were $C_4^1+C_4^2+C_4^3+C_4^4=15$ combinations, represented by the 15 blue circles in Fig.~\ref{fig:unit}. We then connected these circles in the inclusive order, i.e., if two points $A$ and $B$ are on the same curve and $A$ is on the left of $B$, then the electrodes used to obtain $A$ were contained in the electrodes used to obtain $B$ (e.g., $A$ may be obtained from Electrodes 1 and 2, and $B$ from Electrodes 1, 2 and 3, or $B$ from Electrodes 1, 2, and 4). As a result, unlike in Fig.~\ref{fig:ele}, where each dataset has only one curve, in Fig.~\ref{fig:unit} each dataset may have multiple branches leading to the same end-point. To make the curves more distinguishable, we only show the results for 10 datasets in Fig.~\ref{fig:unit}. In general, the classification accuracy increased with the number of units, which is intuitive.

\subsection{4-Class Classification Using Single-Unit and Multi-Unit Spikes}

We calculated the violation rate of the inter spike interval (ISI), which equalled the proportion of ISI shorter than three milliseconds. We defined a single-unit as a well isolated unit when the ISI violation rate is less than 0.02 \cite{Hill2011}. The statistics are shown in Table~\ref{tab:data} ($nU$ denotes the number of units, $nS$ the number of single units, and $nM$ the number of multi-units. Generally, $nU=nS+nM$, except in the third and 33rd datasets where one unit had almost no spikes and hence was removed).

We studied the 4-class classification performance with single-unit (denoted as sSpikes) and multi-unit (denoted as mSpikes) and combined units (single-unit + multi-unit, denoted as Spikes), respectively. For each condition, we performed single-view and multi-view 4-class classification on the 45 datasets for 30 times. The mean and standard deviation of the test classification accuracies are shown in Table~\ref{tab:avg2}. Observe that:
\begin{enumerate}
\item For classification using LFPs only, the features did not change, but the accuracies changed very slightly between three cases. This is because that the partition of training, validation and test sets were different in these three cases.
\item For classification using only spikes, Spikes had better performance than sSpikes and mSpikes. This indicates that Spikes contain more information than sSpikes and mSpikes, which is intuitive.
\item For classification using both LFPs and spikes, LFPs + Spikes had better performance than LFPs+sSpikes and LFPs+mSpikes. This again indicates that Spikes contain more information than sSpikes and mSpikes.
\item Combining LFP and spike features always improved the decoding performance.
\item Our proposed MvBLS achieved the best decoding performance in all conditions.
\end{enumerate}

\begin{table*}[htpb] \centering
\caption{Mean and standard deviation of 4-class classification accuracies (\%) when different classifiers and features were used.} \label{tab:avg2}
\begin{tabular}{c|cccccc}
\hline
               & SVM            & Ridge                   & BLS                     & MvDA           & MvMDA          & MvBLS                   \\ \hline
LFPs           & 40.57$\pm$0.70 & 40.85$\pm$0.70          & \textbf{41.24$\pm$0.62} & $-$            & $-$            & $-$                     \\
Spikes         & 43.49$\pm$0.62 & \textbf{44.21$\pm$0.47} & 43.48$\pm$0.59          & $-$            & $-$            & $-$                     \\
LFPs+Spikes  & 46.12$\pm$0.56 & 46.51$\pm$0.41          & 45.65$\pm$0.52          & 39.49$\pm$0.50 & 36.62$\pm$0.55 & \textbf{47.94$\pm$0.62} \\ \hline
LFPs           & 40.61$\pm$0.60 & 40.72$\pm$0.62          & \textbf{41.10$\pm$0.58} & $-$            & $-$            & $-$                     \\
sSpikes        & 34.73$\pm$0.52 & \textbf{35.93$\pm$0.50} & 35.26$\pm$0.53          & $-$            & $-$            & $-$                     \\
LFPs+sSpikes & 41.46$\pm$0.62 & 41.54$\pm$0.63          & 41.20$\pm$0.58          & 36.41$\pm$0.63 & 32.92$\pm$0.79 & \textbf{43.37$\pm$0.77} \\ \hline
LFPs           & 40.41$\pm$0.61 & 40.66$\pm$0.51          & \textbf{40.96$\pm$0.47} & $-$            & $-$            & $-$                     \\
mSpikes        & 39.35$\pm$0.58 & \textbf{40.06$\pm$0.62} & 39.77$\pm$0.59          & $-$            & $-$            & $-$                     \\
LFPs+mSpikes & 43.92$\pm$0.47 & 44.04$\pm$0.64          & 43.36$\pm$0.51          & 38.64$\pm$0.58 & 35.05$\pm$0.69 & \textbf{45.16$\pm$0.57} \\ \hline
\end{tabular}
\end{table*}

\subsection{Binary Classification Using Single-Unit and Multi-Unit Spikes}

Four-class classification was considered in previous subsections, where we decoded one location out of all four possibilities. This subsection considers binary classification, where we decode one location out of the two available choices. Specifically, based on the classification confidences of the four possible locations, we calculated the binary test accuracies given the two locations available for choice. We performed single-view and multi-view classification on the 45 datasets for 30 times. The mean and standard deviation of binary test classification accuracies are shown in Table~\ref{tab:avg3}.

\begin{table*}[htpb] \centering
\caption{Mean and standard deviation of binary classification accuracies (\%) when different classifiers and features were used.} \label{tab:avg3}
\begin{tabular}{c|cccccc}
\hline
               & SVM                     & Ridge          & BLS                     & MvDA           & MvMDA          & MvBLS                   \\ \hline
LFPs           & 63.49$\pm$0.51          & 62.52$\pm$0.58 & \textbf{63.55$\pm$0.52} & $-$            & $-$            & $-$                     \\
Spikes         & \textbf{67.47$\pm$0.63} & 67.14$\pm$0.66 & 67.25$\pm$0.67          & $-$            & $-$            & $-$                     \\
LFPs+Spikes  & 69.13$\pm$0.56          & 68.45$\pm$0.60 & 68.24$\pm$0.53          & 62.67$\pm$0.66 & 60.84$\pm$0.63 & \textbf{69.51$\pm$0.56} \\ \hline
LFPs           & \textbf{63.89$\pm$0.59} & 62.67$\pm$0.63 & 63.71$\pm$0.62          & $-$            & $-$            & $-$                     \\
sSpikes        & 60.32$\pm$0.73          & 60.15$\pm$0.73 & \textbf{60.39$\pm$0.72} & $-$            & $-$            & $-$                     \\
LFPs+sSpikes & 64.83$\pm$0.68          & 63.92$\pm$0.58 & 64.16$\pm$0.56          & 60.27$\pm$0.68 & 57.57$\pm$0.66 & \textbf{65.71$\pm$0.59} \\ \hline
LFPs           & 63.51$\pm$0.50          & 62.52$\pm$0.58 & \textbf{63.61$\pm$0.50} & $-$            & $-$            & $-$                     \\
mSpikes        & \textbf{64.23$\pm$0.68} & 63.91$\pm$0.45 & 64.09$\pm$0.74          & $-$            & $-$            & $-$                     \\
LFPs+mSpikes & 67.28$\pm$0.53          & 66.39$\pm$0.54 & 66.28$\pm$0.55          & 62.05$\pm$0.60 & 59.35$\pm$0.59 & \textbf{67.58$\pm$0.53} \\ \hline
\end{tabular}
\end{table*}

The binary classification accuracies in Table~\ref{tab:avg3} are significantly above chance (50\%). Intuitively, they are also much higher than their 4-class classification counterparts in Table~\ref{tab:avg2}. However, the observations made in the previous subsection still hold. Specifically, Spikes outperformed sSpikes and mSpikes, LFPs+Spikes outperformed LFPs+sSpikes and LFPs+mSpikes, combining LFP and spike features always improved the decoding performance, and our proposed MvBLS always achieved the best decoding performance.

\subsection{MvBLS Parameter Sensitivity} \label{sect:ParameterSensitivity}

MvBLS has three structural parameters and three nomalization/regularization parameters ($n$, $m$, $k$, $s$, $\lambda_1$ and $\lambda_2$ in Algorithm~\ref{alg:MvBLS1}). It is important to know the sensitivity of MvBLS to them, which will provide valuable guidelines in selecting these parameters in future applications.

\begin{figure}[htpb] \centering
\includegraphics[width=\linewidth,clip]{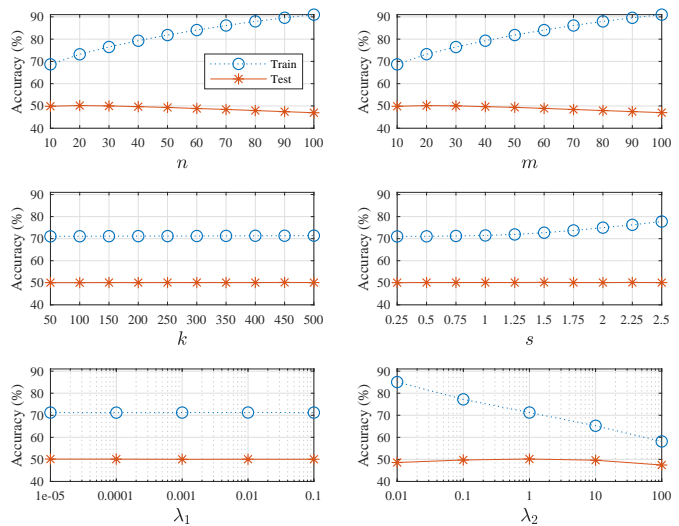}
\caption{MvBLS classification (4-class) accuracy versus its parameters.} \label{fig:MVBLSParam}
\end{figure}

By default $n=15$, $m=15$, $k=300$, $s=0.8$, $\lambda_1=0.001$ and $\lambda_2=1$. When studying the sensitivity of MvBLS to $n$, we fixed $m$, $k$, $s$, $\lambda_1$ and $\lambda_2$ at their default values, and varied $n$ from $10$ to $100$. For each $n$ on each dataset, we trained 30 MvBLSs on 30 different partitions of the dataset (80\% for training and 20\% for test), and recorded the average test accuracy across the 30 runs. Finally we took the average of the 45 datasets, and show the results in the top-left panel of Fig.~\ref{fig:MVBLSParam}. Similarly, we varied $m$ from $10$ to $100$, $k$ from $50$ to $500$, $s$ from $0.25$ to $2.5$, $\lambda_1$ from $10^{-5}$ to $0.1$, and $\lambda_2$ from $0.01$ to $100$, and show the results in Fig.~\ref{fig:MVBLSParam}. Observe that:
\begin{enumerate}
\item As $n$ or $m$ increased, the training accuracy increased quickly, but the test accuracy slightly decreased. This suggests that smaller $n$ and $m$ should be used for better generalization performance, which is beneficial to the computational cost.
\item The training and test accuracies almost did not change with $k$ and $\lambda_1$. So, we can set $k$ to be a small value to save the computational cost, and choose $\lambda_1$ safely in $[0.00001,0.1]$.
\item As $s$ increased, the training accuracy increased slowly, but the test accuracy almost did not change. So, we can choose $s$ safely in $[0.25, 2.5]$.
\item As $\lambda_2$ increased, the training accuracy decreased very quickly, but the test accuracy first increased slightly and then decreased slightly. This suggests that the regularization should not be too small or too large, which is a well-known fact in machine learning.
\end{enumerate}

In general, we conclude that MvBLS is robust to its parameters.

\subsection{Computational Cost}

Next, we compared the computational cost of different algorithms, as in practice a faster algorithm is preferred over a slower one, given similar classification accuracies.

We recorded the mean running time (including training, validation and test time) of 45 sessions for different classifiers, when different features were used. Since this process was repeated 30 times, we obtained 30 mean running time for each algorithm-feature combination. The mean and standard deviation for each combination were computed from these 30 numbers and shown in Table~\ref{tab:avgTime}. The platform was a Linux workstation with Intel Xeon CPU (E5-2699@2.20GHz) and 500GB RAM. SVM was the most efficient single-view learning algorithm, and MvBLS the most efficient multi-view learning algorithm. Particularly, MvBLS was several times faster than the other two state-of-the-art multi-view learning approaches, and it also achieved the best performance. In summary, our proposed MvBLS is both effective and efficient.

It is important to note that identifying the best model parameters in training is time-consuming, because a lot of cross-validations are needed; however, once the optimal model parameters are found, all models can be run very fast in test, which involves mostly matrix operations.

\begin{table*}[htpb] \centering
\caption{Mean and standard deviation of running time (seconds; including cross-validation to identify the best model parameters) when different classifiers and features were used in 4-class classification.} \label{tab:avgTime}
\begin{tabular}{c|cccccc}
\hline
              & SVM                     & Ridge                  & BLS             & MvDA             & MvMDA            & MvBLS                    \\ \hline
LFPs          & 7.72$\pm$0.30           & \textbf{4.40$\pm$0.53} & 37.52$\pm$2.24  & $-$              & $-$              & $-$                      \\
Spikes        & \textbf{19.05$\pm$2.92} & 585.54$\pm$52.87       & 85.77$\pm$5.42  & $-$              & $-$              & $-$                      \\
LFPs+Spikes & \textbf{24.84$\pm$3.19} & 1033.14$\pm$89.47      & 102.40$\pm$6.83 & 748.06$\pm$66.55 & 741.87$\pm$68.08 & \textbf{129.66$\pm$7.93} \\ \hline
\end{tabular}
\end{table*}

To better illustrate the above point, we removed the validation set and compared SVM and MvBLS when both LFPs and spikes were used as features. The box constraint of SVM was set to $10^{-4}$, which was the most frequently chosen value in Section~III-G. The parameters of MvBLS were $n=15$, $m=15$, $k=300$, $s=0.8$, $\lambda_1=0.001$ and $\lambda_2=1$, which were the default values in Section~IV-D. We also compared them with a multi-view deep learning approach, deep canonically correlated autoencoders (DCCAE) \cite{Wang2015}. DCCAE aligned the two views of data in a common subspace and then used an ECOC-SVM classifier for classification. Its parameters were adjusted from the open source code\footnote{https://ttic.uchicago.edu/$\sim$wwang5/dccae.html} by changing the mini-batch size for the correlation and the reconstruction error terms from 800 to 16, and the maximum number of epoches from 14 to 9, and the box constraint of SVM was set to $1$, as they gave higher classification accuracies. We ran the three models on each of the 45 sessions eight times. Each time we randomly partitioned the datasets into 80\% training and 20\% test. The platform was a laptop computer with AMD Ryzen 5 CPU (3550H@2.10GHz) and 16GB RAM, running Windows 10 x64 and Matlab 2020a. The mean and standard deviation of 4-class and binary classification accuracies on the test set, as well as the computing time (including both training and test on one dataset), are shown in Table~\ref{tab:SingleModle}\footnote{The accuracies here were slightly higher than those in Tables~\ref{tab:avg2} and \ref{tab:avg3}, because the training set was larger.}. MvBLS achieved the highest classification accuracies, and was also the fastest.

\begin{table}[htpb] \centering
\caption{Mean and standard deviation of 4-class and binary classification accuracies, as well as the computing time, for different models, when both LFPs and Spikes were used as features.} \label{tab:SingleModle}
\begin{tabular}{c|ccc}
\hline
                      & SVM            & DCCAE            & MvBLS                   \\ \hline
4-class accuracy (\%) & 46.77$\pm$0.05 & 39.20$\pm$0.07   & \textbf{50.35$\pm$0.08} \\
Binary accuracy (\%)  & 69.77$\pm$0.08 & 65.32$\pm$0.08   & \textbf{71.57$\pm$0.04} \\
Computing time (s)    & 5.26$\pm$0.08  & 1409.20$\pm$1.31 & \textbf{1.99$\pm$0.04}  \\ \hline
\end{tabular}
\end{table}

\subsection{Additional MvBLS Approaches}

\begin{figure}[htpb] \centering
\subfigure[]{\label{fig:MvBLS2}     \includegraphics[width=.8\linewidth,clip]{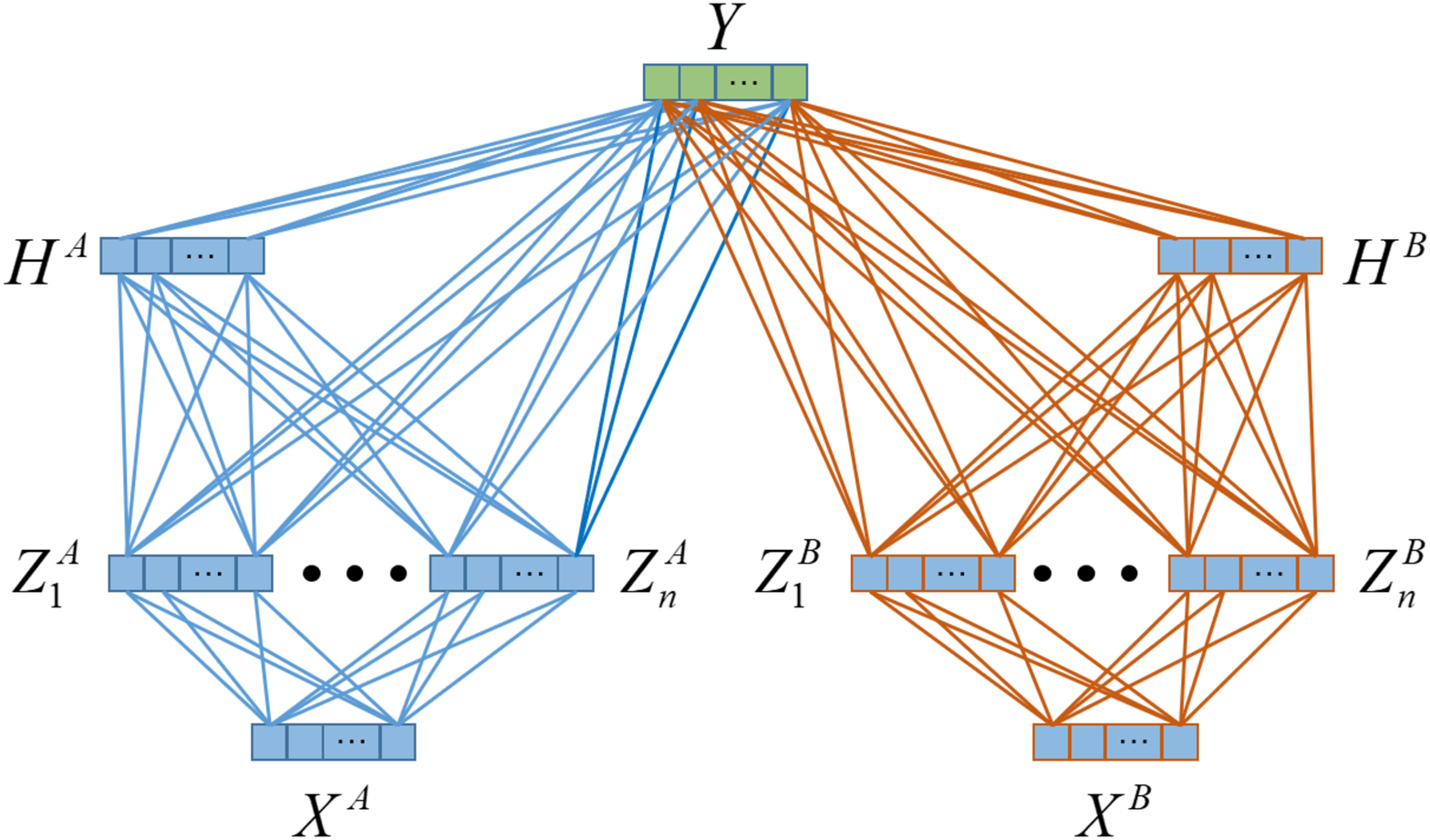}}
\subfigure[]{\label{fig:MvBLS3}     \includegraphics[width=.8\linewidth,clip]{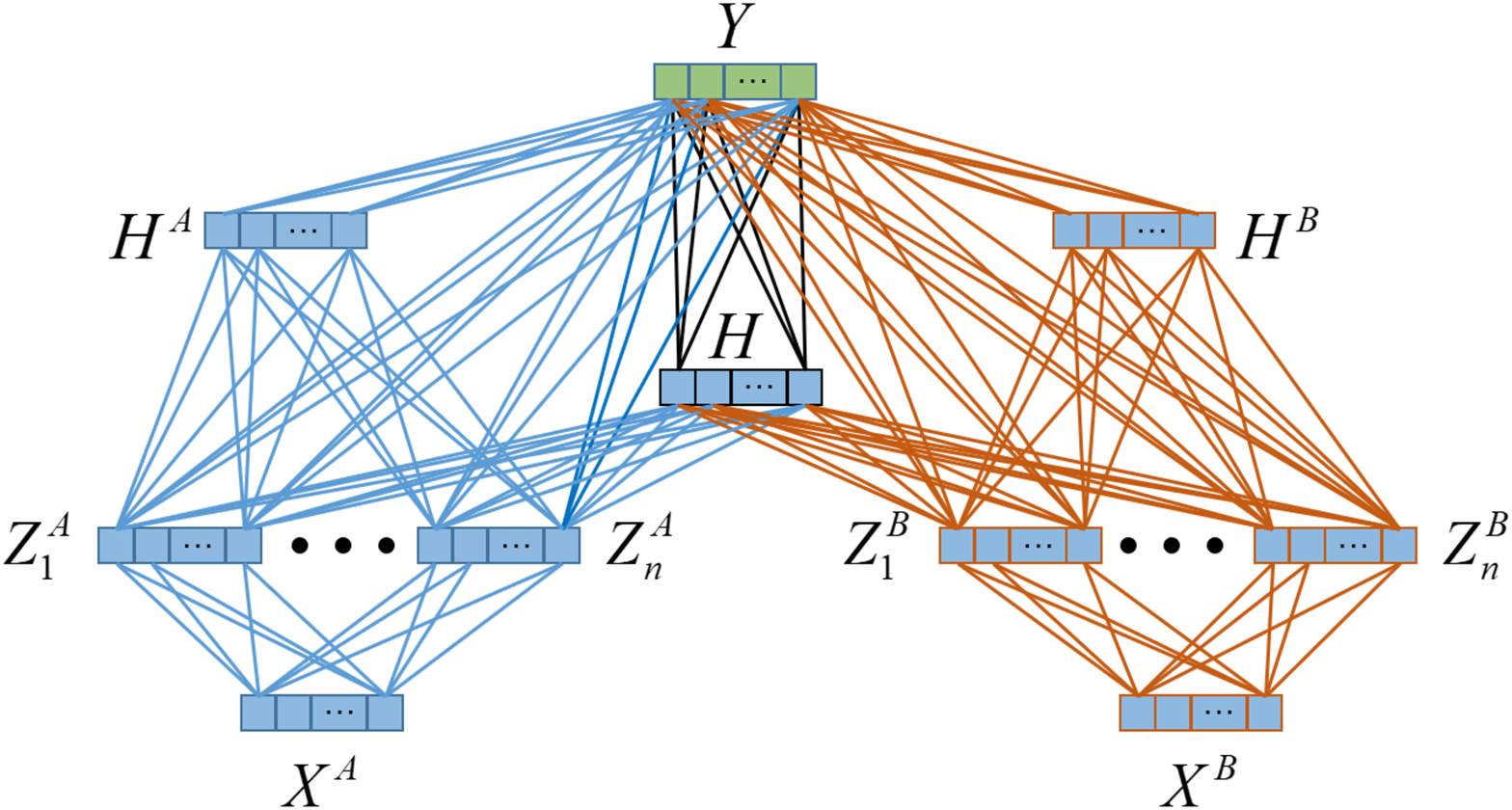}}
\caption{Two additional configurations of MvBLS. (a) MvBLS2; (b) MvBLS3.} \label{fig:MVBLSs}
\end{figure}

In addition to the MvBLS model in Fig.~\ref{fig:MvBLS1}, other MvBLS architectures can also be configured, by constructing the inputs to $\mathbf{Y}$ differently. Two additional configurations are shown in Fig.~\ref{fig:MVBLSs}, and denoted as MvBLS2 and MvBLS3, respectively. Compared with MvBLS in Fig.~\ref{fig:MvBLS1}, MvBLS2 in Fig.~\ref{fig:MvBLS2} first constructs enhancement nodes $\mathbf{H}^A$ from $\mathbf{Z}^A$ and $\mathbf{H}^B$ from $\mathbf{Z}^B$, and then feeds all of them into $\mathbf{Y}$; so, it has more nodes and weights than MvBLS. Compared with MvBLS2, MvBLS3 in Fig.~\ref{fig:MvBLS3} further constructs enhancement nodes $\mathbf{H}$ from $\mathbf{Z}^A$ and $\mathbf{Z}^B$, and then feeds $\mathbf{H}$, $\mathbf{H}^A$, $\mathbf{H}^B$, $\mathbf{Z}^A$ and $\mathbf{Z}^B$ into $\mathbf{Y}$. So, MvBLS3 has even more nodes and weights than MvBLS2.

\begin{figure*}[htpb] \centering
\includegraphics[width=\linewidth,clip]{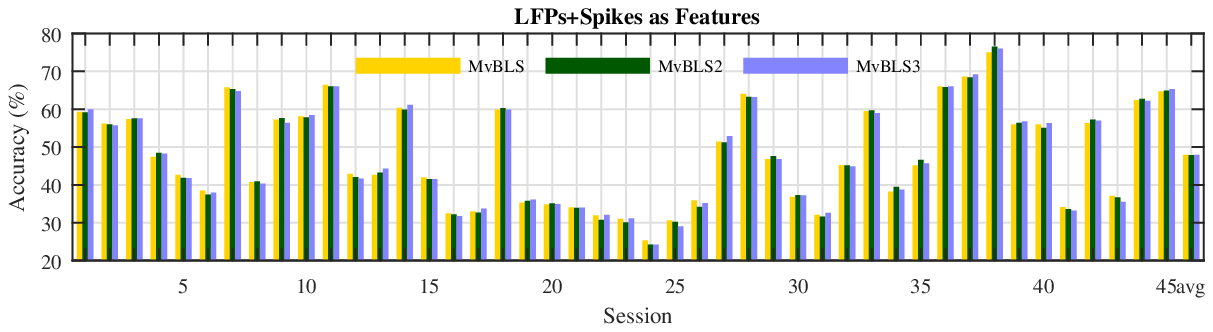}
\caption{4-class classification accuracies of the three different MvBLSs.} \label{fig:MVBLScomp}
\end{figure*}

The 4-class classification performances of MvBLS, MvBLS2 and MvBLS3 in the 45 sessions are shown in Fig.~\ref{fig:MVBLScomp}. Their classification accuracies were almost identical, which is interesting, considering that MvBLS2 and MvBLS3 had more parameters and connections. Indeed, non-parametric multiple comparisons showed that there was no statistically significant difference between any two of them. Since MvBLS has much simpler configuration and is easier to train, it is preferred in our application.

\subsection{Related Work} \label{sect:related}

Both LFPs and spikes contain information about the monkeys' oculomotor decision, and there has been independent research on both. Spikes are high-pass filtered neural signals, which can be used to decode high-performance movement control signals \cite{Humphrey1970,Santhanam2006A}. However, since spikes often deteriorate as electrodes degrade over time, more stable LFPs, which are low-pass filtered neural signals, are used in long term BMIs \cite{Andersen2004Selecting,Rickert2005Encoding,Zhuang2010Decoding,Bansal2011Relationships,Flint2012Accurate,Flint2013Long}.

Because LFPs and spikes can be recorded from the same electrodes \cite{Monosov2008}, and convey complementary information \cite{Belitski2008Low,Buzsaki2012,Chen2018,Einevoll2013}, a natural approach is to combine them for more accurate decoding \cite{Bokil2006method,Perge2014Reliability,Stavisky2015high,Ibayashi2018Decoding,Bansal2011Decoding,Hsieh2019}.

Bokil \emph{et al.} \cite{Bokil2006method} trained two macaque monkeys to perform a memory-saccade task and collected LFPs and spikes from the lateral intraparietal area. Two-dimensional Fourier transforms were performed to extract the features. Saccade prediction was achieved by maximizing the log-likelihood function of the observed neural activity. This approach was novel in that it did not use trial start time or other trial-related timing information. However, the performance degraded when switching from the preferred-or-anti-preferred binary classification to four-direction and eight-direction classifications.

Bansal \emph{et al.} \cite{Bansal2011Decoding} trained two male macaque monkeys to perform reach-and-grasp tasks in three dimensions, and collected 192-channel LFPs and spikes from primary and ventral premotor areas. Linear Gaussian state-space representation and Kalman filter were then used to decode the reach-and-grasp kinematics. The decoding was first conducted for each channel, then about 30 channels were iteratively chosen based on the decoding performance (the Pearson correlation coefficient between the measured and the reconstructed kinematics). This approach required a large number of channels to be chosen from, which may not available in many human and non-human primate studies, including ours.

Hsieh \emph{et al.} \cite{Hsieh2019} trained one adult rhesus macaque to perform a center-out-and-back task and collected 137-channel LFPs and spikes from dorsal premotor cortex and ventral premotor cortex of both hemispheres. They then developed a multi-scale encoding model, a multi-scale adaptive learning algorithm, and a multi-scale filter for decoding the millisecond time-scale of spikes and slower LFPs. This approach solved a trajectory regression problem, whereas we focused on oculomotor decision classification.

Most studies, except \cite{Bansal2011Decoding} and \cite{Hsieh2019} introduced above have not shown significant improvements in decoding performance, compared with using LFPs or spikes alone. Our research has shown that sophisticated machine learning approaches like MvBLS can better fuse LFPs and spikes, and hence achieve significant decoding performance improvements.

\section{Conclusion}

Multi-view learning is beneficial for decoding oculomotor decisions using medial frontal neural signals from non-human primates. This is because these simultaneously recorded neural signals comprise both low-frequency LFPs and high-frequency spikes, which can be treated as two views of the brain state. In this paper, we have extended single-view BLS to MvBLS, and validated its performance in monkey oculomotor decision decoding from medial frontal LFPs and spikes. We demonstrated that primate medial frontal LFPs and spikes do contain complementary information about the oculomotor decisions, and that the proposed MvBLS is a more effective approach to use these two types of information in decoding the decision, than several classical and state-of-the-art single-view and multi-view learning approaches. Moreover, we showed that MvBLS is fast, and robust to its parameters. Therefore, we expect that MvBLS will find broader applications in other primate brain state decoding tasks, and beyond.


\end{document}